\documentclass[aps,prd,eqsecnum,floats,twocolumn,showpacs]{revtex4}

\usepackage{graphicx}
\usepackage{bm}

\newcommand{\third}{{\scriptstyle\frac{1}{3}}}
\newcommand{\half}{{\scriptstyle\frac{1}{2}}}
\newcommand{\fourth}{{\scriptstyle\frac{1}{4}}}
\newcommand{\eqref}[1]{(\ref{#1})}
\newcommand{\khat}{\hat k}
\newcommand{\zhat}{\hat z}
\newcommand{\ahat}{\hat a}
\newcommand{\bhat}{\hat b}
\newcommand{\chat}{\hat c}
\newcommand{\dhat}{\hat d}
\newcommand{\ehat}{\hat e}
\newcommand{\kbar}{\bar k}
\newcommand{\zbar}{\bar z}
\newcommand{\abar}{\bar a}
\newcommand{\bbar}{\bar b}
\newcommand{\cbar}{\bar c}
\newcommand{\dbar}{\bar d}
\newcommand{\ebar}{\bar e}

\begin{document}

\title{Energy Norms and the Stability of the Einstein Evolution Equations}

\author{Lee Lindblom and Mark A. Scheel}

\affiliation{Theoretical Astrophysics 130-33, California Institute of
Technology, Pasadena, CA 91125}

\date{\today}

\begin{abstract}
The Einstein evolution equations may be written in a variety of
equivalent analytical forms, but numerical solutions of these
different formulations display a wide range of growth rates for
constraint violations.  For symmetric hyperbolic formulations of the
equations, an exact expression for the growth rate is derived using an
energy norm. This expression agrees with the growth rate determined by
numerical solution of the equations. An approximate method for
estimating the growth rate is also derived. This estimate can be
evaluated algebraically from the initial data, and is shown to exhibit
qualitatively the same dependence as the numerically-determined rate
on the parameters that specify the formulation of the equations.  This
simple rate estimate therefore provides a useful tool for finding the
most well-behaved forms of the evolution equations.  
\end{abstract}

\pacs{04.25.Dm, 04.20.Ex, 02.60.Cb, 02.30.Mv}

\maketitle

\section{Introduction}
\label{s:introduction}
It is well known that the Einstein equations may be written in a
variety of forms%
~\cite{ADM,york79,shibata95,baumgarte99,%
fischer_marsden79,bona_masso92,frittelli_reula94,bona_masso95b,%
choquet_york95,abrahams_etal95,mvp96,frittelli_reula96,friedrich96,%
estabrook_etal97,Iriondo1997,anderson_etal98,Bonilla1998,Yoneda1999,%
anderson_york99,Frittelli1999,Alcubierre1999,Hern1999,%
Friedrich_Rendall2000,Yoneda2000,%
Shinkai2000,Kidder2001,Yoneda2001,Yoneda2002a,%
Alvi2002,Sarbach2002,Yoneda2002b,Laguna2002}.
In recent years a growing body of work has
documented the fact that these different formulations, while
equivalent analytically, 
have significantly different stability properties when
used for unconstrained~\footnote{By unconstrained evolution we mean that
the evolution equations are used to advance the fields in time,
but the constraint equations are used only to set initial data
and to check the accuracy of the results.}
numerical evolutions%
~\cite{scheel_etal97b,baumgarte99,Shinkai2000,%
Kidder2001,Laguna2002,Alcubierre2000b,Calabrese2002}.
The most important of these differences is the behavior of
non-physical solutions of the evolution equations, which often grow
exponentially and eventually dominate the desired physical
solutions. These non-physical solutions could be solutions of the
evolution equations that violate the constraints
(``constraint-violating instabilities'') or solutions that satisfy the
constraints but represent some ill-behaved coordinate transformation
(``gauge instabilities'').  In many cases it is the rapid exponential
growth of these non-physical solutions, rather than numerical issues,
that appear to be the key factor that limits our ability to run
numerical simulations of black holes for long
times~\cite{scheel_etal97b,Alcubierre2000a,Kidder2001,Knapp2002}.
For lack of a better term, we refer to these non-physical
solutions as ``instabilities'' (because they are unstable, {\it
i.e.,\/} exponentially-growing, solutions of the evolution equations),
but keep in mind that they are neither numerical instabilities nor do
they represent physics.

In this paper we explore the use of the energy norm (which can be
introduced for any symmetric hyperbolic form of the evolution
equations) to study these instabilities.  We derive an exact
expression for the growth rate in terms of the energy norm, and verify
that the rate determined in this way agrees with the growth rate of
the constraint violations determined numerically.  We also derive an
approximate expression for this growth rate that can be evaluated
algebraically from the initial data for the evolution equations.  We
explore the accuracy of this approximation by comparing it with
numerically-determined growth rates for solutions of a family of
symmetric hyperbolic evolution equations.

In order to compare the analytical expressions for the growth rates
derived here with the results of numerical computations, it is
necessary to select some particular family of evolution equations with
which to make the comparisons.  Here we focus our attention on the
12-parameter family of first-order evolution equations introduced by
Kidder, Scheel, and Teukolsky (KST)~\cite{Kidder2001}. This family of
equations has been shown to be strongly hyperbolic when certain
inequalities are satisfied by the 12 parameters; however, our
expressions for the instability growth rates apply only to {\it
symmetric\/} hyperbolic systems of equations. Therefore we must extend
the analysis of the KST equations by explicitly constructing the
symmetrizer (or metric on the space of fields) that makes the
equations symmetric hyperbolic.  We show that such a symmetrizer (in
fact a four-parameter family of such symmetrizers) can be constructed
for an open subset of the KST equations having only physical
characteristic speeds.

We compare numerical evolutions of the symmetric hyperbolic subset of
the KST equations with our analytical expressions (both exact and
approximate) for the growth rates of the instabilities.  We make these
comparisons using two sets of initial data for the evolution
equations: flat space in Rindler coordinates~\cite{Wald}, and the
Schwarzschild geometry in Painlev\'e-Gullstrand
coordinates~\cite{painleve21,gullstrand22,Martel2000}.  We find that
our exact analytic expression for the growth rate of the instability
agrees with the actual growth rate of the constraints in (both fully
nonlinear and linearized) numerical simulations.  This agreement
provides further evidence that the constraint-violating instabilities
are real features of the evolution equations and not an artifact of
using a poor numerical algorithm.  In addition, the approximate
analytical expressions for the growth rates derived here are shown to
have good qualitative agreement with the numerically-determined rates.
This approximation therefore provides a useful tool for finding
more well-behaved formulations of the equations.  Furthermore, the
growth rate of the instability is shown here to depend in a
non-trivial way on the exact ``background'' solution as well as on the
particular formulation of the equations.  Hence, unfortunately, it
seems likely that it will never be possible to find a unique ``most
stable'' form of the equations for the evolution of all initial data.

\section{Energy Norms and Rate Estimates}
\label{s:energynorms}

We limit our study here to formulations of the Einstein evolution
equations that can be expressed as first-order systems
\begin{equation}
\partial _t u^\alpha + A ^{k\alpha}{}_\beta \partial _k u^\beta =
F ^\alpha.
\label{e:firstordereqs}
\end{equation}
Here $u ^\alpha$ is the collection of dynamical fields, $A
^{k\alpha}{}_\beta$ and $F ^\alpha$ are (generally complicated)
functions of $u ^\alpha$, and $\partial _t$ and $\partial _k$ are the
partial derivatives with respect to the time $t$ and the spatial
coordinates $x ^k$ respectively.  (We use Greek
indices to label the dynamical fields and Latin indices to label
spatial components of tensors.)  Systems of equations of the
form~(\ref{e:firstordereqs}) are called weakly hyperbolic~\cite{Kreiss1989} 
if $n _k A^{k\alpha}{}_\beta$ has all real eigenvalues for all spatial
one-forms $n_k$, 
and strongly hyperbolic if in addition $n_k A
^{k\alpha}{}_\beta$ has a complete set of eigenvectors for all $n_k$.
There exists a large literature devoted to a
variety of representations of the Einstein evolution equations that
satisfy these conditions%
~\cite{fischer_marsden79,bona_masso92,frittelli_reula94,bona_masso95b,%
choquet_york95,abrahams_etal95,mvp96,frittelli_reula96,friedrich96,%
estabrook_etal97,Iriondo1997,anderson_etal98,Bonilla1998,Yoneda1999,%
anderson_york99,Frittelli1999,Alcubierre1999,Hern1999,%
Friedrich_Rendall2000,Yoneda2000,Shinkai2000,Kidder2001,%
Yoneda2001,Yoneda2002a,%
Alvi2002,Sarbach2002}.  
In particular, the 12-parameter KST system of equations that we use for
our numerical comparisons is of this form.

In order to construct an energy norm, first-order systems such as
Eq.~\eqref{e:firstordereqs} must have an additional structure: a
``symmetrizer'' $S_{\alpha\beta}$.  First-order systems of evolution
equations are called symmetric hyperbolic~\cite{Kreiss1989} (or
symmetrizable hyperbolic) if there exists a symmetrizer which serves
as a metric on the space of fields.  Such a symmetrizer must be
symmetric and positive definite (i.e. $S _{\alpha\beta}u ^\alpha u
^\beta > 0,\,\,\, \forall\,\, u ^\alpha \neq 0$); in addition, it must
symmetrize the matrices $A ^{k\alpha}{}_\beta$: $S _{\alpha\mu} A
^{k\mu}{}_\beta\equiv A ^k_{\alpha\beta}= A ^k_{\beta\alpha},\,\,\,
\forall\,\,{\scriptstyle k}$.  In this paper we limit our discussion
to symmetric hyperbolic formulations.  Note that symmetric hyperbolic
systems are automatically strongly hyperbolic, because symmetric
matrices $n _k A ^k_{\alpha\beta}$ always have real eigenvalues with a
complete set of eigenvectors.  But the converse is not true: strongly
hyperbolic systems need not be symmetric hyperbolic (except in one
spatial dimension).  In Sec.~\ref{s:kstequations} we construct
symmetrizers for (an open subset of) the KST equations.

Let us turn now to the question of the stability of the evolution
equations.  To do this we consider solutions to the equations that are
close (as defined by the metric $S _{\alpha\beta}$) to an exact
``background'' solution $u^\alpha_e$~\footnote{We will typically
consider $u^\alpha_e$ that also satisfy the initial data constraints,
but this is not essential.}. Note that $u^\alpha_e$ may be
time-dependent.  We define $\delta u ^\alpha=u ^\alpha-u ^\alpha_e$ to
be the deviation of the solution $u ^\alpha$ from this given
background solution.  The evolution of $\delta u ^\alpha$ is
determined by the linearized evolution equations:
\begin{equation}
\partial_t \delta u^\alpha 
+ A ^{k\alpha}{}_{\beta}\partial _k\delta u ^\beta
= F ^{\alpha}{}_{\beta}\delta u ^\beta.
\label{e:lineareq}
\end{equation}
Here $A ^{k\alpha}{}_{\beta}$ and $F ^{\alpha}{}_{\beta}$ may depend
on $u ^\alpha_e$ but not on $\delta u ^\alpha$.  We illustrate in
Fig.~\ref{f:fig1} below that the constraint-violating instabilities
occur in the solutions to these linearized evolution equations as well
as in the solutions to the full non-linear equations.

\subsection{Energy Evolution}

For any symmetric hyperbolic system of evolution equations, we may
define a natural ``energy density'' and
``energy-flux''~\cite{courant-hilbert,Gustafsson1995} associated with
$\delta u^\alpha$:
\begin{eqnarray}
\delta E &=& S _{\alpha\beta}\delta u ^\alpha \delta u ^\beta,\\
\delta E ^k &=& A ^{k}_{\alpha\beta}\delta u ^\alpha 
\delta u ^\beta.
\end{eqnarray}
It follows immediately from the linearized evolution equations
Eq.~\eqref{e:lineareq} that this energy density evolves as follows,
\begin{equation}
\label{eq:dtenergy}
\partial _t\,\delta E + \nabla _k\,\delta E^k = C_{\alpha\beta}\delta u ^\alpha
\delta u ^\beta,\label{e:energyeq}
\end{equation}
where $\nabla _k$ is the spatial covariant derivative associated
with the (background) three-metric $g_{ij}$, $C_{\alpha\beta}$
is given by
\begin{eqnarray}
C _{\alpha\beta}&=&2S _{\mu(\alpha}F ^\mu{}_{\beta)} + \partial _t 
S _{\alpha\beta}\nonumber\\
&&+(\sqrt{g}) ^{-1}\partial _k\bigl(\sqrt{g}
A ^{k}_{\alpha\beta}\bigr),
\end{eqnarray}
and $g=\det g_{ij}$ is the determinant of the (background)
spatial metric. Note that $C _{\alpha\beta}$, which serves as
the source (or sink) for the energy in~(\ref{e:energyeq}), depends on
$u ^\alpha_e$ but not on $\delta u ^\alpha$.

\subsection{Exact Expression for the Growth Rate}
\label{sec:exact-expr-growth}
Next we explore the possibility of using this energy to measure
and to estimate the growth rate of instabilities.
Define the growth rate $1/\tau$ of the energy norm to be
\begin{equation}
\frac{1}{\tau}=\frac{\partial _{\,t} ||\delta E||}{2||\delta E||},
\end{equation}
where the energy norm $||\delta E||$ is defined by
\begin{equation}
||\delta E|| = \int\delta E\sqrt{g} \, d ^{\,3}x.
\end{equation}
Integrating Eq.~\eqref{e:energyeq} over a $t$=constant
surface, we obtain the following general expression for the growth rate of
the energy norm:
\begin{equation}
\frac{1}{\tau}=\frac{1}{2||\delta E||}
\int \left(C_{\alpha\beta}\delta u^\alpha\delta u^\beta 
-\nabla_n\delta E^n\right)\sqrt{g}\, d^{\,3}x.
\label{e:taudef}
\end{equation}
We note that Eq.~\eqref{e:taudef} is an identity for any solution of
the equations.  The rate $1/\tau$ becomes independent of time when
$\delta u ^\alpha$ grows exponentially: $\delta u ^\alpha\propto e
^{t/\tau}$.

\begin{figure}
\begin{center}
\includegraphics[width=2.5in]{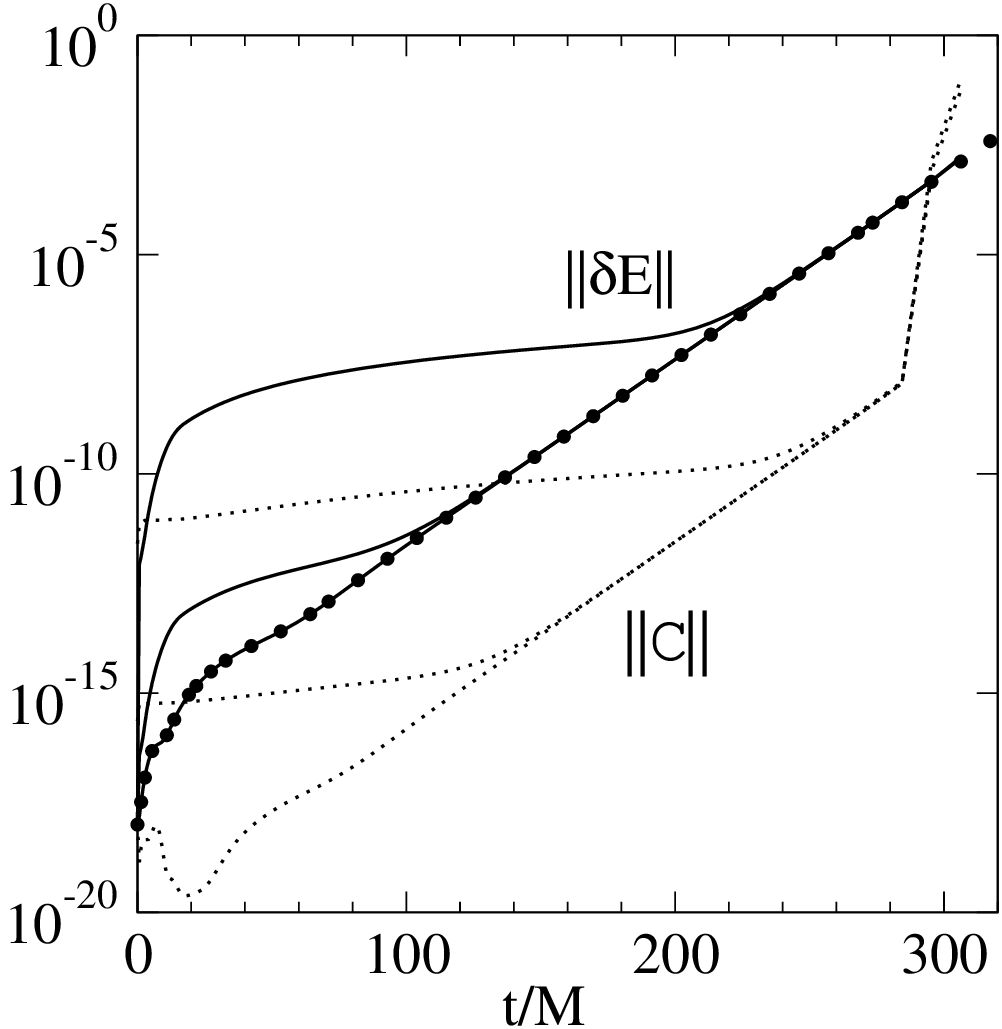}
\end{center}
\caption{Energy norm $||\delta E||$ and
constraints $||{\cal C}||$ (per unit volume) for evolutions of
perturbed Schwarzschild initial data using three spectral resolutions.
Solid curves are $||\delta E||$ from the full non-linear evolution
code, and points are from the linearized code.  Dotted curves are
$||{\cal C}||$ from the non-linear code.}
\label{f:fig1}
\end{figure}

Figure~\ref{f:fig1} illustrates the equivalence between the energy
norm measure and the standard measures of the growth rate of the
constraint-violating instability.   Plotted
are results from 3D nonlinear numerical evolutions of perturbed
Schwarzschild initial data using a particular formulation of the
Einstein evolution equations~\footnote{ KST System
III~\cite{Kidder2001} with $\eta=4/33$ (or $\gamma=-16$) and
$\hat{z}=-1/4$.}.  The solid curves show the evolution of the energy
norm $||\delta E||$ while the dotted curves show the evolution of the
norm of the constraint violations,
\begin{equation}
||{\cal C}|| \equiv \int \bigl( {\cal C} ^2 + 
{\cal C} _i  {\cal C} ^i\bigr)\sqrt{g}\, d ^{\,3}x,
\end{equation}
where $ {\cal C}$ represents the Hamiltonian and $ {\cal
C} _i$ the momentum constraints~\footnote{The actual evolution
equations studied here involve not only these physical constraints,
but a large number of additional constraints as well.  We illustrate in
Sec.~\ref{sec:rindler-spacetime} that all the constraints grow at the
same rate.}.  The larger
points plotted in Fig.~\ref{f:fig1} show the energy norm computed for
a numerical solution of the {\it linearized\/} evolution equations,
indicating that the constraint-violating instabilities occur even in
the linearized theory.

Figure~\ref{f:fig1} clearly illustrates that the constraint violations
$||{\cal C}||$ grow at the same rate as the energy norm
$||\delta E||$ (until the very end of the simulation when non-linear
effects become significant).  This equality between the growth rates
is exact for any constraint-violating solution having the form $\delta
u^\alpha(t,\vec x) =e^{t/\tau}\delta u^\alpha(0,\vec x)$.  Our
numerical solution approaches this form asymptotically.  The
numerically-determined slopes of these curves, $1/\tau({\delta
E_L})=0.0489$ (from the linear evolution code), $1/\tau({\delta
E_{NL}})=0.0489$ (from the non-linear code) and $1/\tau({\cal
C})=0.0490$ (also from the non-linear code), are also in good
agreement with the growth rate determined from the integral expression
in Eq.~\eqref{e:taudef}: $1/\tau({\scriptscriptstyle \int})=0.0489$.
This agreement shows that the numerical solutions satisfy the identity
in Eq.~\eqref{e:taudef}.  This provides further strong evidence that
the constraint-violating instabilities seen here are real solutions to
the evolution equations, rather than arising from purely numerical
problems associated with the discrete representation of the solution
or the time-evolution algorithm.

The computational domain, boundary conditions, initial data, and other
details of the numerical evolutions shown in Fig.~\ref{f:fig1} are the
same as described later in Section~\ref{sec:schw-spac}.  To choose
gauge conditions we set the shift and the densitized lapse equal to
their analytic values for all time.  Each nonlinear evolution in
Fig.~\ref{f:fig1} is shown for three different spectral resolutions,
18x8x15, 24x8x15, and 32x8x15 (where the notation
$N_r$x$N_\theta$x$N_\phi$ represents the number of spectral
collocation points in the $r$, $\theta$, and $\phi$ directions),
demonstrating the asymptotic convergence of these results.  The
results for the same three resolutions using the linear evolution code
are indistinguishable from each other in Fig.~\ref{f:fig1}, so only
one resolution is plotted. These linearized results are also essentially
indistinguishable (until very late times) from the highest resolution
non-linear results.

\subsection{Approximate Expression for the Growth Rates}

Although Eq.~\eqref{e:taudef} is an identity, it does not provide a
particularly useful way to determine $1/\tau$.  Its use requires
the full knowledge of the spatial structure of the unstable solution
$\delta u ^\alpha$, and this can be determined only by solving the
equations.  Our goal is to obtain a reasonable estimate of $1/\tau$
without having to solve the evolution equations.  

We first note that if $\delta E^k \,n_k \ge 0$ at the boundaries
(where $n_k$ is the outward-directed normal one-form at
the boundary), one can
integrate~(\ref{e:taudef}) by parts and obtain
\begin{equation}
\frac{1}{\tau} \le \frac{1}{2||\delta E||} 
\int C_{\alpha\beta}\delta u^\alpha\delta u^\beta \, \sqrt{g} d^{\,3}x.
\end{equation}
Therefore if $\lambda_{\rm max}$ is the largest eigenvalue $\lambda$ of
the equation 
$0=(C_{\alpha\beta} - \lambda S_{\alpha\beta})\delta u^\beta$,
then
\begin{equation}
\label{eq:growthrateupperbound}
\frac{1}{\tau} \le \frac{\lambda_{\rm max}}{2},
\end{equation}
or equivalently,
\begin{equation}
\label{eq:energynormupperbound}
||\delta E|| \le C ||\delta E_{(t=0)}|| e^{\lambda_{\rm max} t}
\end{equation}
for some constant $C$.  This argument is often used~\cite{Gustafsson1995} to
show that symmetric hyperbolic systems have well-posed initial value
problems, {\it i.e.,\/} that symmetric hyperbolic systems have growth
rates that are bounded by exponentials.  Because our numerical
simulations use boundary conditions that satisfy $\delta E^kn_k\ge 0$
(incoming characteristic fields are zero), we could
use~(\ref{eq:growthrateupperbound}) to estimate the growth rate.
Unfortunately, we find that the upper bound provided
by~(\ref{eq:growthrateupperbound}) is typically far greater than the
actual growth rate, and therefore an estimate based on this bound is
not very useful.

Therefore we take a different approach.  Without any prior knowledge
of the structure of the actual unstable solution, the simplest choice
is to assume that the spatial gradients are given approximately by
$\partial_n\delta u^\alpha\approx k_n\delta u^\alpha$, where $k_n$ is
a ``wavevector'' that characterizes the direction and magnitude of the
gradient of $\delta u^\alpha$.  (Since the actual solutions
that are responsible for the instabilities in the cases we have
studied do seem to have a characteristic lengthscale, typically the
mass of the black hole or some other physically distinguished scale,
this approximation should not be too bad.)  In this case the
expression in Eq.~\eqref{e:taudef} for $1/\tau$ simplifies to
\begin{equation}
\frac{1}{\tau}=\frac{1}{2||\delta E||}
\int \bar C _{\alpha\beta}\delta u ^\alpha\delta u ^\beta 
\sqrt{g}\,d^{\,3}x,\label{e:tauapproxdef}
\end{equation}
where $\bar C _{\alpha\beta}$ is given by
\begin{equation} \bar C _{\alpha\beta}=2S _{\mu(\alpha}F ^\mu{}_{\beta)} +
\partial _t S _{\alpha\beta}-2 k_n A^n_{\alpha\beta}.
\end{equation}

Next we limit the $\delta u ^\alpha$ to the subspace of field vectors
that satisfy the boundary conditions.  We do this formally for these
$\delta u ^\alpha$ by introducing the projection operator $P
^\alpha{}_\beta$.  This projection annihilates vectors that violate
the boundary conditions, and leaves those vectors that satisfy all the
boundary conditions unchanged~\footnote{This projection operator is
not unique, however, in the cases that we study here there is a
obvious simple choice.}.  Using this projection we re-write
Eq.~\eqref{e:tauapproxdef} as
\begin{equation}
\frac{1}{\tau}=\frac{1}{2||\delta \tilde E||}
\int \bar C _{\alpha\beta}P ^\alpha{}_\mu P ^\beta{}_\nu
\delta u ^\mu\delta u ^\nu 
\sqrt{g}\,d^{\,3}x,\label{e:tauapproxdef2}
\end{equation}
where $\delta \tilde E =S _{\alpha\beta}P ^\alpha{}_\mu P ^\beta{}_\nu
\delta u ^\mu\delta u ^\nu$.  We expect that the fastest growing
solution to the evolution equation will be the one driven most
strongly by these ``source'' terms on the right side of
Eq.~\eqref{e:tauapproxdef2}.  Thus we approximate (roughly) this most
unstable solution as the eigenvector $\delta e^\alpha$ of $\bar
C_{\alpha\beta}P ^\alpha{}_\mu P ^\beta{}_\nu$ having the largest
eigenvalue:
\begin{equation}
{\bar \lambda}_{\max} S_{\alpha\beta}P ^\alpha{}_\mu P
^\beta{}_\nu\delta e^\nu = \bar C_{\alpha\beta}P ^\alpha{}_\mu P
^\beta{}_\nu\delta e^\nu.
\end{equation}
The integrals in Eq.~\eqref{e:tauapproxdef2} are easily evaluated for
this eigenvector $\delta e^\alpha$, giving the following approximate
expression for the growth rate,
\begin{equation}
\frac{1}{\tau}\approx\frac{1}{2\int \delta \tilde E \sqrt{g}\,d^{\,3}x}
\int{\bar \lambda}_{\max} \delta \tilde E
\sqrt{g}\,d^{\,3}x=\frac{{\bar \lambda}_{\max}}{2}.\label{e:tauestimate}
\end{equation}
In this approximation then the growth rate of this most unstable
solution is half the maximum eigenvalue of
$\bar C_{\alpha\beta}P ^\alpha{}_\mu P ^\beta{}_\nu$.   We test the
accuracy of this approximation in Sec.~\ref{s:growthrates} by comparing its
predictions with the results of numerical solutions to the evolution
equations.

\section{Einstein Evolution Equations}
\label{s:kstequations}

In this Section we introduce the particular formulations of the
Einstein evolution equations that we study numerically to make
comparisons with the growth rate estimates derived in
Sec.~\ref{s:energynorms}.  The rather general 12-parameter family of
formulations introduced by Kidder, Scheel, and Teukolsky
(KST)~\cite{Kidder2001} is ideal for these purposes.  In this Section
we review these formulations and derive expressions for the
symmetrizer $S _{\alpha\beta}$ and energy density $\delta E$ (when they exist)
that are needed for our growth rate estimates.  This Section contains
a very brief review of the derivation and basic properties of the KST
equations using the notation of this paper, followed by a rather more
detailed and technical derivation of the needed symmetrizer and energy
norms.  Readers more interested in our numerical tests of the growth
rate estimates might prefer to skip ahead to that material in
Sec.~\ref{s:growthrates}.

\subsection{Summary of the KST Equations}

The KST formulation of the Einstein evolution equations begins with
the standard ADM~\cite{ADM} equations (discussed in detail
in~\cite{york79}) written as a first-order system for the
``fundamental'' dynamical variables: $u ^\alpha_0=\{g _{ij}, K _{ij},
D_{kij}\}$, where $g _{ij}$ is the spatial metric, $K _{ij}$ is the
extrinsic curvature, and $D _{kij}\equiv\half\partial _k g _{ij}$.  We
express these standard ADM equations in the (somewhat abstract) form:
\begin{eqnarray}
\partial _{\,t} g{}_{ij}&=&N ^k\partial _k g _{ij}+2g _{k(i}\partial _{j)}
N ^k-2NK _{ij}  ,\label{e:kstg}\\
\partial _{\,t} K _{ij} &=& \cdots,\label{e:kstp}\\
\partial _{\,t} D _{kij} &=& \cdots,\label{e:kstm}
\end{eqnarray}
where $N$ and $N ^k$ are the lapse and shift respectively.  The
$\,\,\cdots\,\,$ on the right sides of Eqs.~\eqref{e:kstp} and
\eqref{e:kstm} stand for the standard terms that appear in the
first-order form of the ADM equations, which are given explicitly (up
to a slight change in notation~\footnote{Our $D_{kij}$ differs from
the quantity $d_{kij}$ used by KST by a factor of two:
$d_{kij}=2D_{kij}$.}) in Eqs.~(2.14) and (2.24) of
KST~\cite{Kidder2001}.  The 12-parameter extension of these equations
proposed by KST splits naturally into two parts: the first part has 5
{\it dynamical} parameters (represented by Greek letters $\gamma$,
$\zeta$, $\eta$, $\chi$, and $\sigma$) that influence the dynamics
(e.g. including the characteristic speeds) of the system in a
fundamental way, and the second part has 7 {\it kinematical}
parameters (represented by Latin letters $\zhat$, $\khat$, $\ahat$,
$\bhat$, $\chat$, $\dhat$, and $\ehat$) that merely re-define the
dynamical fields.

The 5-parameter family of {\it dynamical} modifications of these
equations is obtained {\it a)} by adding a 4-parameter family of
multiples of the constraints to the fundamental ADM equations, and
{\it b)} by assuming that a certain densitized lapse function (which
depends on one additional parameter) rather than the lapse itself, is
a fixed function on spacetime.  The first modification is obtained by
adding multiples of the constraints to the right sides of
Eqs.~\eqref{e:kstp} and \eqref{e:kstm}:
\begin{eqnarray}
\partial _{\,t} K _{ij} &=& \cdots\,\,+\gamma N g _{ij} {\cal C}
+ \zeta N g ^{ab} {\cal C} _{a(ij)b},\label{e:kstp1}\\
\partial _{\,t} D _{kij} &=& \cdots\,\,+\half \eta N g _{k(i} {\cal C}
_{j)}+\half\chi N g _{ij} {\cal C} _k
,\label{e:kstm1}
\end{eqnarray}
where $\gamma$, $\zeta$, $\eta$, and $\chi$ are constants, and the
constraints (for the vacuum case) are defined by
\begin{eqnarray}
{\cal C}&=&\half[{}^{(3)}R-K _{ij} K ^{ij}+K^2],\label{e:ch} \\
{\cal C} _i &=& \nabla _j K ^j{}_i - \nabla _i K,\\
{\cal C} _{kij} &=& \partial _k g _{ij} - D _{kij},\\
{\cal C} _{klij}&=&\partial _k D _{lij}-\partial _l D _{kij}\label{e:ct}.
\end{eqnarray}
The second modification comes by assuming that the densitized lapse $Q$,
defined by
\begin{equation}
Q=\log(Ng ^{-\sigma}),
\end{equation}
(rather than the lapse itself) is a fixed function on spacetime, where
$g=\det(g _{ij})$ and $\sigma$ is a constant.  With these
modifications the extended ADM equations become a 5-parameter family
of evolution equations for the fundamental fields $u ^\alpha_0$.  These
equations can be written as
\begin{equation}
\partial _{\,t} u ^\alpha_0 + A _0^{k\alpha}{}_{\beta}
\partial _{\,k} u ^\beta_0 = F ^\alpha_0.\label{e:firstorderpde}
\end{equation}
The quantities $A ^{k}_0{}^\alpha{}_\beta$ and $F ^\alpha_0$ are
functions of the fields $u ^\alpha_0$ and the parameters $\gamma$,
$\zeta$, $\eta$, $\chi$, and $\sigma$.  We give
explicit expressions for the $A ^{k}_0{}^\alpha{}_\beta$ in
Sec.~\ref{s:symmetric} and the Appendix.

The 12-parameter KST family of representations of the Einstein
evolution equations is completed by adding a 7-parameter family of
{\it kinematical} transformations of the dynamical fields to the
fundamental representations given in Eq.~\eqref{e:firstorderpde}.
These transformations replace $K _{ij}$ and $D _{kij}$ by $P _{ij}$
and $M _{kij}$ according to the expressions:
\begin{eqnarray}
&&P _{ij}=K _{ij}+\zhat\, g _{ij}g ^{ab} K _{ab},\label{e:pdef}\\
&&M _{kij}=\bigl[\khat\, \delta ^a_k\delta ^b_i\delta ^c_j
+\ehat\, \delta ^a_{(i}\delta ^b_{j)}\delta ^c_k
+\ahat\, g _{ij} g ^{bc}\delta ^a_k
+\bhat\, g _{ij} g ^{ab}\delta ^c_k\nonumber\\
&&\quad\qquad\quad
+\chat\, g _{k(i} \delta ^a_{j)}g ^{bc}
+\dhat\, g _{k(i} \delta ^c_{j)}g ^{ab}
\bigr]D _{abc}.\label{e:mdef}
\end{eqnarray}
While these kinematical transformations may seem like ``trivial''
re-parameterizations of the theory, numerical results (see
e.g. Sec.~\ref{sec:schw-spac} below) have shown that these transformations
can have a significant effect on the stability of the equations.  The
general transformation defined in Eqs.~\eqref{e:pdef} and
\eqref{e:mdef} is a linear transformation from the basic
dynamical fields $u^\alpha_0$ to the new set of dynamical fields $u
^\alpha=\{g _{ij}, P _{ij}, M _{kij}\}$:
\begin{equation}
u ^\alpha=T ^\alpha{}_\beta u ^\beta_0,\label{e:utrans}
\end{equation}
where the transformation $T ^\alpha{}_\beta$ depends on the
kinematical parameters and the metric $g _{ij}$.  The special case
with $\khat=1$ and $\zhat=\ahat=\bhat=\chat=\dhat=\ehat=0$ corresponds
to the identity transformation $u^\alpha=u^\alpha_0$.  The explicit
representations of the inverse transformation $(T
^{-1})^{\,}{}^\alpha{}_\beta$ are given in the Appendix.

The evolution equations for the new transformed dynamical fields $u
^\alpha$ are obtained by multiplying Eq.~\eqref{e:firstorderpde} by $T
^\alpha{}_\beta$.  The result (after some straightforward
manipulation) has the same general form as
Eq.~\eqref{e:firstorderpde},
\begin{equation}
\partial _{\,t} u ^\alpha + A ^{k\alpha}{}_{\beta}
\partial _{\,k} u ^\beta = F ^\alpha,\label{e:firstorderpde1}
\end{equation}
with $A ^k{}^\alpha{}_\beta$ and $F ^\alpha$ given by: 
\begin{eqnarray} A ^{k\alpha}{}_\beta&=&T ^\alpha{}_\mu A ^{k\mu}_0{}_\nu
(T ^{-1})^{\,\nu}{}_\beta,\label{e:atransform}\\ 
F ^\alpha &=& T ^\alpha{}_\beta F ^\beta_0 + V ^{\alpha ij}\partial _t
g _{ij} + 2 A ^k{}^\alpha{}_\beta V ^{\beta ij}D _{kij},\qquad\quad
\label{e:ftransform} 
\end{eqnarray}
where $V ^{\alpha ij}$ is defined by
\begin{equation}
V ^{\alpha ij}= \frac{\partial T ^\alpha{}_\mu}{\partial g _{ij}}
(T ^{-1}){}^\mu{}_\nu u ^\nu.\label{e:vaij}
\end{equation}
All of the terms on the right side of Eq.~\eqref{e:ftransform}, except
the term containing $\partial _t g _{ij}$, depend on the $u ^\alpha$
(or $u ^\alpha_0= T ^{-1}{}^\alpha{}_\beta u^\beta$) and not its
derivatives.  In the remaining term, the quantity $\partial_t g_{ij}$
is to be replaced by the right-hand side of Eq.~\eqref{e:kstg}; this
introduces the spatial derivatives $\partial_k g_{ij}$, which are to
be replaced by $2D_{kij}$.  Thus in the end $F ^\alpha$ in
Eq.~\eqref{e:ftransform} is simply a function of $u ^\alpha$ as
required.

The simple transformation~\eqref{e:atransform} that relates the
matrix $A ^{k\mu}_0{}_\nu$ of the fundamental representation of the
equations with $A ^{k\mu}{}_\nu$ ensures that the characteristic
speeds of the theory are independent of the kinematical parameters:
\begin{equation}
0=\det\left(-v\delta^\alpha{}_\beta
+n _kA ^{k\alpha}{}_\beta\right)=\det\left(-v\delta^\alpha{}_\beta
+n _kA ^{k\alpha}_0{}_\beta\right).
\end{equation}
These characteristic speeds (relative to the hypersurface orthogonal
observers) are also independent of direction $n _k$ in general
relativity.  Thus the hyperbolicity of these formulations can depend
only on the dynamical parameters $\gamma$, $\zeta$, $\eta$, $\chi$,
and $\sigma$.  One can show that these characteristic speeds (relative
to the hypersurface orthogonal observers) are $\{0,\pm 1,\pm v_1,\pm
v_2,\pm v_3\}$, where
\begin{eqnarray}
v_1^2&=&2\sigma,\label{e:v1}\\
v_2^2&=&{\scriptstyle \frac{1}{8}}\left(\eta-4\eta\sigma-2\chi-12\sigma\chi
-3\eta\zeta\right),\\
v_3^2&=&\half\left(2+4\gamma-\eta-2\gamma\eta+2\chi+4\gamma\chi-\eta\zeta
\right).\qquad\label{e:v3}
\end{eqnarray}

In much of the analysis that follows, we will restrict attention to
the subset of these KST equations where the characteristic speeds have
only the physical values $\{0,\pm 1\}$.  This requires
$v_1^2=v_2^2=v_3^2=1$ if the theory is also to be strongly hyperbolic
(see KST~\cite{Kidder2001}).  Thus our primary focus will be the 9-parameter
family of equations in which the parameters $\sigma$, $\eta$, and
$\chi$ are fixed by the conditions,
\begin{eqnarray}
\sigma&=&\half,\label{e:sigma}\\
\eta &=& {-8\over 5 + 7\zeta + 10 \gamma + 6 \gamma\zeta},\label{e:eta}\\
\chi & = & -{4 + 4\zeta + 10 \gamma + 6\gamma\zeta 
\over 5 + 7\zeta + 10 \gamma + 6 \gamma\zeta}\label{e:chi},
\end{eqnarray}
that are needed to ensure $v_1=v_2=v_3=1$.
The parameters $\gamma$ and $\zeta$ are arbitrary so long as
$5+7\zeta+10\gamma+6\gamma\zeta\neq0$.  Thus the curve
$\gamma=-(7\zeta+5)/(6\zeta+10)$ is forbidden, but $\gamma$ and
$\zeta$ are otherwise unconstrained~\footnote{We note that this
representation in terms of $\gamma$ and $\zeta$ encompasses both
cases of the representation in terms of $\eta$ and $\zeta$ given
by KST in their Eqs.~(2.38) and (2.39)~\cite{Kidder2001}.}.

\subsection{Symmetric Hyperbolicity of KST}
\label{s:symmetric}

A first order system such as Eq.~\eqref{e:firstorderpde1} is called
symmetric hyperbolic if there exists a positive definite
\emph{symmetrizer} $S{}_{\alpha\beta}$ such that $A{}^n_{\alpha\beta}
\equiv S{}_{\alpha\gamma}A{}^{n\gamma}{}_\beta$ is symmetric in the
indices $\alpha$ and $\beta$,
$A{}^n_{\alpha\beta}=A{}^n_{\beta\alpha}$, for all field
configurations.  We assume here that $S{}_{\alpha\beta}$ depends only
on the spatial metric $g{}_{ij}$~\footnote{While this assumption is
not essential, such a symmetrizer must exist if there is any positive
definite symmetrizer that depends only on the dynamical fields $u
^\alpha$.  In a neighborhood of flat space only $g{}_{ij}$ is non-zero
so any symmetrizer in this neighborhood must depend on $g _{ij}$.}.
It is convenient to represent the symmetrizer as a quadratic form
\begin{equation}
dS^2=S{}_{\alpha\beta} du {}^{\alpha} du {}^\beta,
\end{equation}
where $du {}^\alpha=\{dg {}_{ij}, dP _{ij},dM _{kij}\}$ denotes the
standard basis of co-vectors on the space of dynamical fields.  The
most general symmetric quadratic form on the space of dynamical fields
(which depends only on the metric $g{}_{ij}$) is given by
\begin{eqnarray}
&&dS^2=A_1\,dG^2 +B{}_1\, dP^2+2D_1\,dGdP
\nonumber\\ &&\quad
+A_2\,g{}^{ik}g{}^{jl}d\tilde
g {}_{ij} d\tilde g {}_{kl} +B{}_2\, g{}^{ik}g{}^{jl}d \tilde
P {}_{ij}d\tilde P {}_{kl}\nonumber\\&&\quad
+2D_2\,g{}^{ik}g{}^{jl}d\tilde 
g _{ij}d\tilde P _{kl}
+C{}_1g{}^{kl}g{}^{ia}g{}^{jb} d\tilde M {}_{(kij)} d\tilde
M {}^{}_{(lab)}\nonumber\\ &&\quad+C{}_2 g{}^{kl}g{}^{ia}g{}^{jb}
\bigl[d\tilde M {}_{kij}-d\tilde M {}_{(kij)}\bigr] \bigl[d\tilde
M {}^{}_{lab}-d\tilde M {}^{}_{(lab)}\bigr] \nonumber\\ &&\quad+
C{}_3\,g{}^{ij} dM {}^{1}_{i}dM {}^{1}_{j} +C{}_4\,g {}^{ij}
dM {}^{2}_{i} dM{}^{2}_{j}
\nonumber\\ &&\quad+2C{}_5\,g{}^{ij}
dM {}^{1}_{i}dM {}^{2}_{j}.
\label{e:sab} 
\end{eqnarray}
Here $dG$ and $dP$ are the traces of $dg _{ij}$ and $dP {}_{ij}$
respectively, and $d\tilde g {}_{ij}$ and $d\tilde P {}_{ij}$ are their
trace-free parts.  The two
traces of $dM {}_{kij}$ are defined by
\begin{eqnarray}
dM {}^{1}_{i}&\equiv&dM {}_{ijk}g{}^{jk}\\ 
dM {}^{2}_{i}&\equiv&d M {}_{kij}g{}^{jk},
\end{eqnarray}
and its trace-free part, $d\tilde M{}_{kij}$, is
\begin{eqnarray}
d\tilde M {}_{kij} \equiv dM {}_{kij}&+& {\scriptstyle \frac{1}{5}}
\bigl[dM {}^{1}_{(i}g {}_{j)k}-2\,dM {}^{1}_{k} g {}_{ij}\nonumber\\
      &&\quad+ dM {}^{2}_{k} g {}_{ij}-3\, dM {}^{2}_{(i}g {}_{j)k}\bigr].
\qquad 
\end{eqnarray}
This quadratic form, Eq.~\eqref{e:sab}, is positive definite iff
$\{A_1,A_2, B_1, B_2, C_1,C_2,C_3,C_4\}$ are all positive and
$C_5^2 < C_3 C_4$, $D_1^2 < A_1B_1$, and $D_2^2< A_2B_2$.  (The signs
of $C_5$, $D_1$ and $D_2$ are irrelevant.)

The question now is whether the constants $A{}_A$, $B{}_A$, $C{}_A$
and $D_A$ can be chosen to make the $A {}^n_{\alpha\beta}$ symmetric
in $\alpha$ and $\beta$.  In order to answer this we need explicit
expressions for the matrices $A ^{n\alpha}{}_{\beta}$.  Quite
generally these matrices are determined for these equations by a set
of 12 constants $\mu_A$ and $\nu_A$, which in turn are determined by
the 12 parameters of the KST formulations.  (We give the explicit
expressions for $\mu_A$ and $\nu_A$ in terms of the KST parameters in
the Appendix.)  The equations that define the $A
^{n\,\alpha}{}_{\beta}$ in terms of these constants are:
\begin{eqnarray} \partial{}_t g{}_{ij}&\simeq& N ^n\partial _n g
_{ij},\label{ea:kstg}\\ \partial{}_t P {}_{ij}&\simeq& N ^n\partial _n
K _{ij}-N\Bigl[ \mu{}_1 g{}^{nb}\delta{}^c{}_i\delta{}^d{}_j +\mu{}_2
g{}^{nd}\delta{}^b{}_{(i}\delta{}^c{}_{j)} \nonumber\\ &&+\mu{}_3
g{}^{bc}\delta{}^n{}_{(i}\delta{}^d{}_{j)} +\mu{}_4
g{}^{cd}\delta{}^n{}_{(i}\delta{}^b{}_{j)}\nonumber\\ &&+\mu{}_5
g{}^{nd}g{}^{bc}g{}_{ij} +\mu{}_6 g{}^{nb}g{}^{cd}g{}_{ij}\Bigr]
\partial{}_k M {}_{bcd},\label{ea:kstp}\\ \partial{}_t M{}_{kij}
&\simeq& N ^n \partial _n M _{kij} -N\Bigl[ \nu{}_1
\delta{}^n{}_k\delta{}^b{}_i\delta{}^c{}_j
+\nu{}_2\delta{}^n{}_{(i}\delta{}^b{}_{j)}\delta{}^c{}_k\nonumber\\
&&+\nu{}_3 g{}^{nb}g{}_{k(i}\delta{}^c{}_{j)} + \nu{}_4
g{}^{nb}g{}_{ij}\delta{}^c{}_{k}\nonumber\\ &&+ \nu{}_5
g{}^{bc}g{}_{k(i}\delta{}^n{}_{j)} + \nu{}_6
g{}^{bc}g{}_{ij}\delta{}^n{}_{k}\Bigr]\partial{}_n P {}_{bc},
\label{ea:kstm}
\end{eqnarray}
where $\simeq$ means that only the principal parts of the equations
have been represented explicitly ($\partial _t u ^\alpha + A
 ^{n\,\alpha}{}_\beta\partial _k u ^\beta\simeq 0$).

We now evaluate $A^n_{\alpha\beta}=S _{\alpha\mu} A ^{n\,\mu}{}_\beta$
and $A^n_{\beta\alpha}=S _{\beta\mu} A ^{n\,\mu}{}_\alpha$ using the
expressions in Eqs.~\eqref{e:sab} through \eqref{ea:kstm}.  After
lengthy algebraic manipulations, we find that $A^n_{\alpha\beta}$ is
symmetric iff $D_1=D_2=0$ and the following constraints are satisfied
by the constants $B{}_A$ and $C{}_A$:
\begin{eqnarray}
0&=&B_2 (\mu_1 + \mu_2) - C_1 (\nu_1 + \nu_2),\label{e:cb}\\ 
0&=&B_2 (2\mu_1-\mu_2) - C_2(2\nu_1-\nu_2),\qquad\\ 
0&=&B_1 (3 \mu_1 + 3 \mu_4 + 9 \mu_6) 
\nonumber \\ &&
- C_3 (3 \nu_1 + \nu_2 + \nu_3 + 3 \nu_4 + 3 \nu_5 + 9 \nu_6) 
\nonumber \\ &&
- C_5 (\nu_1 + 2 \nu_2 + 2 \nu_3 + \nu_4 + 6 \nu_5 + 3 \nu_6),\\ 
0&=&B_1 (3 \mu_2 + 3 \mu_3 + 9 \mu_5) 
\nonumber \\ &&
- C_5 (3 \nu_1 + \nu_2 + \nu_3 + 3 \nu_4 + 3 \nu_5 + 9 \nu_6) 
\nonumber \\ &&
- C_4 (\nu_1 + 2 \nu_2 + 2 \nu_3 + \nu_4 + 6 \nu_5 + 3 \nu_6),\\ 
0&=&B_2 (-2 \mu_1 + 3 \mu_2 + 10 \mu_4) 
- C_3 (10 \nu_2 + 10 \nu_3 + 30 \nu_4) 
\nonumber \\ &&
- C_5 (10 \nu_1 + 5 \nu_2 + 20 \nu_3 + 10 \nu_4),\\ 
0&=&B_2 (~~6 \mu_1 + ~~\mu_2 + 10 \mu_3) 
- C_5 (10 \nu_2 + 10 \nu_3 + 30 \nu_4) 
\nonumber \\ &&
- C_4 (10 \nu_1 + 5 \nu_2 + 20 \nu_3 + 10 \nu_4).
\label{e:ce}
\end{eqnarray}
This is a system of six linear equations for the seven parameters
$B{}_A$ and $C{}_A$.  Thus we expect there to exist solution(s) to these
equations for almost all choices of the $\mu_A$ and $\nu_A$.  However,
there is no guarantee that such solutions will satisfy the positivity
requirements needed to ensure that $S_{\alpha\beta}$ is positive
definite.

We divide into two parts the question of determining when solutions to
Eqs.~\eqref{e:cb} through \eqref{e:ce} exist that satisfy the appropriate
positivity conditions: first the question of when a positive definite
symmetrizer, $S^0_{\alpha\beta}$, exists for the subset of the KST
equations whose dynamical fields are the fundamental fields
$u^\alpha_0$, and second the question of when this fundamental
symmetrizer can be extended to a symmetrizer for the full 12-parameter
set of KST equations.  We consider the second question first.  Assume
that for a given set of dynamical parameters there
exists a positive definite $S^0_{\alpha\beta}$ such that
$S^0_{\alpha\gamma}A^{n\gamma}_0{}_\beta=
S^0_{\beta\gamma}A^{n\gamma}_0{}_\alpha$.  Now define
$S_{\alpha\beta}$:
\begin{equation}
S_{\alpha\beta}=T^{-1\,\mu}{}_\alpha S^0_{\mu\nu}
T^{-1\,\nu}{}_\beta.\label{e:stransform}
\end{equation}
One can verify, using Eq.~\eqref{e:atransform}, that this
$S_{\alpha\beta}$ symmetrizes $A^{n\alpha}{}_\beta$.  Further it
follows, using Eq.~\eqref{e:utrans}, that this $S_{\alpha\beta}$ is
positive definite, 
\begin{equation}
S_{\alpha\beta}u^\alpha u^\beta = S^0_{\alpha\beta}u^\alpha_0u^\beta_0>0,
\label{e:s0trans}
\end{equation}
since $S^0_{\alpha\beta}$ is assumed to be positive definite.  In the
Appendix we give explicit expressions for the constants $B_A$ and
$C_A$ that define $S _{\alpha\beta}$ in terms of the constants $B^0_A$
and $C^0_A$ that define $S^0_{\alpha\beta}$ and the parameters that
define the transformation $T^{-1}{}^\alpha{}_\beta$.

Now we return to the first, and more difficult, question: when does
there exist a positive definite symmetrizer $S^0_{\alpha\beta}$ for
the subset of the KST equations whose dynamical fields are the
fundamental fields $u^\alpha_0$?  At the present time we have not
solved this problem completely.  Rather, we restrict our attention to
an interesting (perhaps the most interesting) subset of these KST
equations in which the characteristic speeds $v_1$, $v_2$ and $v_3$
are all the speed of light: $v_1=v_2=v_3=1$.  The restrictions that
these conditions place on the dynamical parameters are given in
Eqs.~\eqref{e:sigma} through \eqref{e:chi}.  Each of these systems is
strongly hyperbolic.

One can now evaluate the $\mu_A$ and $\nu_A$ appropriate for this
subset of KST equations using Eqs.~\eqref{e:mu10} through
\eqref{e:nu60} along with Eqs.~\eqref{e:sigma} through \eqref{e:chi}.
Substituting these into Eqs.~\eqref{e:cb} through \eqref{e:ce} gives
the symmetrization conditions for these equations.  These conditions
are degenerate in this case, reducing to only five independent
equations.  Solving these five symmetrization equations for
the $C_A^0$ in terms of the $B_A^0$ gives
\begin{eqnarray} C_1^0&=&-\zeta B_2^0,\label{e:c10}\\
C_2^0&=&\half(3+\zeta) B_2^0,\\ C_3^0&=&9 (1+\gamma)^2B_1^0 +
{3(\zeta-5)^2\over 10(5+3\zeta)}B_2^0,\label{e:c30}\\ C_4^0&=&
(2+3\gamma)^2 B_1^0 + {(9\zeta-5)^2\over 30(5+3\zeta)}B_2^0,\\
C_5^0&=& - 3(1+\gamma)(2+3\gamma)B_1^0\nonumber\\
&&\qquad-{(\zeta-5)(9\zeta-5)\over 10(5+3\zeta)}B_2^0.\label{e:c50}
\end{eqnarray}
These equations guarantee that $\{C_1^0,C_2^0,C_3^0,C_4^0\}$ are positive
for any positive $B_1^0$ and $B_2^0$ if and only if 
\begin{equation}
0 > \zeta > -{\scriptstyle \frac{5}{3}}.
\label{e:zetalimits}
\end{equation}
The only remaining condition needed to establish symmetric
hyperbolicity is to ensure that $C_3^0C_4^0-(C_5^0)^2>0$.  Using
Eqs.~\eqref{e:c30} through \eqref{e:c50} it follows that
\begin{equation}
C_3^0C_4^0-(C_5^0)^2 = {3(5+7\zeta+10\gamma+6\gamma\zeta)^2\over
10(5+3\zeta)}B_1^0B_2^0.
\end{equation}
Thus the right side is positive whenever $B_1^0$ and $B_2^0$ are
positive iff $\zeta>-{\scriptstyle\frac{5}{3}}$ and
$5+7\zeta+10\gamma+6\gamma\zeta\neq0$.  The first of these conditions
along with Eq.~\eqref{e:c10} demonstrates that
Eq.~\eqref{e:zetalimits} is the necessary and sufficient constraint on
the parameters $\{\zeta,\gamma\}$ to ensure symmetric hyperbolicity.
The second of these conditions was also required to ensure that the
parameters $\eta$ and $\chi$ in Eqs.~\eqref{e:eta} and \eqref{e:chi}
are finite, so it does not represent a new restriction.  

Thus a large open set of this two-parameter family of the fundamental
KST representations of the Einstein evolution equations is symmetric
hyperbolic.  And perhaps even more surprising, the complimentary
subset of these strongly hyperbolic equations (i.e. when
$\zeta>0$ or $\zeta<-{\scriptstyle \frac{5}{3}}$) is \emph{not}
symmetric hyperbolic~\footnote{The common belief that strongly
hyperbolic systems are also symmetric hyperbolic is true only for
equations on spacetimes with a single spatial dimension.}.  Further,
the extension of this two-parameter family via
Eq.~\eqref{e:stransform} produces a nine-parameter family of strongly
hyperbolic representations the Einstein equations.  A large open
subset of this nine-parameter family is symmetric hyperbolic
(i.e. those that are extensions of the symmetric hyperbolic
fundamental representations), while its compliment is not symmetric
hyperbolic.

The construction used here to build a symmetrizer $S_{\alpha\beta}$
for the KST equations has succeeded unexpectedly well.  We found not
just a single symmetrizer, but in fact a four-parameter family of such
symmetrizers.  Using the expressions for the $C_A$ from
Eqs.~\eqref{e:c10} through \eqref{e:c50}, we see that the symmetrizer
$S^0_{\alpha\beta}$ is a sum of terms that depend linearly on each of
the four parameters $\{A_1,A_2,B^0_1,B^0_2\}$.  Thus we may
write this symmetrizer in the form:
\begin{equation}
S^0_{\alpha\beta}=A_1S^{10}_{\alpha\beta}
+A_2S^{20}_{\alpha\beta}
+B^0_1S^{30}_{\alpha\beta}
+B^0_2S^{40}_{\alpha\beta},\label{e:subsymm}
\end{equation}
where the ``sub-symmetrizers'' $S^{A0}_{\alpha\beta}$ represent a set
of positive definite (under the conditions needed for symmetric
hyperbolicity established above) metrics on mutually orthogonal
subspaces in the space of fields $u ^\alpha$.  The expressions for
these sub-symmetrizers are
\begin{eqnarray}
\lefteqn{S^{10}_{\alpha\beta}du ^\alpha du ^\beta
=dG^2,}\hfill\label{e:s10}\\
\lefteqn{S^{20}_{\alpha\beta}du ^{\alpha}du ^\beta
= g^{ij}g^{kl} d\tilde g _{ij}d\tilde g _{kl},}\hfill\\
\lefteqn{S^{30}_{\alpha\beta}du ^{\alpha}du ^\beta
=dP^2
+g ^{ij}\bigl[3(1+\gamma)dM^{1}_i-(2+3\gamma)dM^{2}_i\bigr]}\hfill\nonumber\\
&&\qquad\qquad\qquad\times
\bigl[3(1+\gamma)dM^{1}_j-(2+3\gamma)dM^{2}_j\bigr],\quad \\
\lefteqn{S^{40}_{\alpha\beta}du^{\alpha}du^\beta
=g^{ik}g^{jl}d\tilde P _{ij}d\tilde P _{kl}
-\zeta g^{kl}g^{ia}g^{jb} d\tilde M _{(kij)}\tilde dM _{(lab)}}
\hfill\nonumber\\
&&\quad+\bigl[30(5+3\zeta)\bigr]^{-1}g^{ij}
\bigl[3(\zeta-5)dM ^{1}_i-(9\zeta-5)dM^{2}_i\bigr]\nonumber\\
&&\qquad\qquad\qquad\times\bigl[3(\zeta-5)dM^{1}_j-(9\zeta-5)dM^{2}_j\bigr]
\nonumber\\
&&\quad+\third(3+\zeta)g^{kl}g^{ia}g^{jb}
\bigl[d\tilde M _{kij}-d\tilde M _{(kij)}\bigr]\nonumber\\
&&\qquad\qquad\qquad\times
\bigl[d\tilde M _{lab}-d\tilde M _{(lab)}\bigr].
\label{e:s40}\end{eqnarray}
The dimensions of the corresponding sub-spaces are $\{1, 5, 4, 20\}$.
Just as the fundamental symmetrizer $S^0_{\alpha\beta}$ is related to
the more general symmetrizer $S_{\alpha\beta}$ by the transformation
given in Eq.~\eqref{e:stransform}, so the fundamental sub-symmetrizers
are related to the general sub-symmetrizers by,
\begin{equation}
S^A_{\alpha\beta}=T^{-1\,\mu}{}_\alpha S^{A0}_{\mu\nu}
T^{-1\,\nu}{}_\beta.\label{e:sAtransform}
\end{equation}
We note that this transformation leaves the first two sub-symmetrizers
unchanged: $S^1_{\alpha\beta}=S^{10}_{\alpha\beta}$ and 
$S^2_{\alpha\beta}=S^{20}_{\alpha\beta}$.

One interesting subset of the nine-parameter family of KST equations
studied here is the two-parameter ``generalized
Einstein-Christoffel'' system studied extensively by
KST~\cite{Kidder2001}.  In the language used here this two-parameter
family is defined by
\begin{eqnarray}
\zeta&=&-\khat=-1,\label{e:sys3.1}\\
\ahat&=&\gamma+3\zhat+3\zhat\gamma,\\
\bhat&=&-\gamma-2\zhat-3\gamma\zhat,\\
\chat&=&-\dhat=2,\\
\ehat&=&0.\label{e:sys3.6}
\end{eqnarray}
Substituting these parameter values into Eqs.~\eqref{e:c10} through
\eqref{e:c50} and \eqref{e:b1} through \eqref{e:c5}, we find the
constants that determine the symmetrizer to be
\begin{eqnarray}
B^0_1&=&(1+3\zhat)^{2} B_1,\\
B^0_2&=&C_1=C_2=B_2,\\
C_3&=&B_1+{\scriptstyle \frac{1}{15}}B_2,\\
C_4&=&{\scriptstyle \frac{3}{5}} B_2,\\
C_5&=&-{\scriptstyle \frac{1}{5}} B_2,
\end{eqnarray}
where $B_1$ and $B_2$ are arbitrary positive constants.  For this
case the sub-symmetrizers $S^3_{\alpha\beta}$ and $S^4_{\alpha\beta}$
have the simple forms:
\begin{eqnarray}
&&S^{3}_{\alpha\beta}du ^{\alpha}du ^\beta
=dP^2+g^{ij}dM^{1}_idM^{1}_j,\hfill\\
&&S^{4}_{\alpha\beta}du ^{\alpha}du ^\beta
=
g^{kl}g^{ia}g^{jb}dM _{kij}dM _{lab}
-\third g^{ij}dM ^{1}_idM^{1}_j\hfill\nonumber\\
&&\qquad\qquad\qquad+g^{ik}g^{jl}dP _{ij}dP _{kl}-\third dP^2.
\end{eqnarray}
We also point out that the symmetrizer becomes particularly simple in
this generalized Einstein-Christoffel case by taking $A_1=B_1=\third$ and
$A_2=B_2=1$:
\begin{eqnarray}
dS^2=S_{\alpha\beta}du ^{\alpha}du ^\beta&=&
g^{ik}g^{jl}dg _{ij}dg _{kl} +g^{ik}g^{jl}dP _{ij}dP _{kl}\nonumber\\
&& + 
g^{kl}g^{ia}g^{jb}dM _{kij}dM _{lab}.\label{e:euclid}
\end{eqnarray}
This represents a kind of ``Euclidean'' metric on the space of fields,
in which the symmetrizer is just the sum of squares of the components of
the dynamical fields.

\subsection{The KST Energy Norms}
\label{s:kstenergynorms}
The symmetrizers for the KST equations may be written as {\it
arbitrary} positive linear combinations of the sub-symmetrizers as in
Eq.~\eqref{e:subsymm}.  Since the equation for $C_{\alpha\beta}$, and
hence the equation for the evolution of the energy is linear in
$S_{\alpha\beta}$, it follows that there are in fact four independent
energy ``sub-norms'' for the KST equations.  Each is defined using the
corresponding sub-symmetrizer:
\begin{eqnarray}
\delta E_A &=& S^A_{\alpha\beta}\delta u^\alpha \delta u^\beta,\\
\delta E^n_A &=& S^A_{\alpha\mu}A^{n\mu}{}_{\beta}\delta u^\alpha 
\delta u^\beta.
\end{eqnarray}
It follows that these energies each satisfy evolution
equations analogous to Eq.~\eqref{e:energyeq}:
\begin{equation}
\partial_t\delta E_A + \nabla_n\delta E^n_A = C^A_{\alpha\beta}\delta u^\alpha
\delta u^\beta,
\end{equation}
where $C^A_{\alpha\beta}$ is given by
\begin{eqnarray}
C^A_{\alpha\beta}&=&2S^A_{\mu(\alpha}F^\mu{}_{\beta)} + \partial_t 
S^A_{\alpha\beta}\nonumber\\
&&+(\sqrt{g})^{-1}\partial_n
\bigl(\sqrt{g}S^A_{\alpha\mu}A^{n\mu}{}_\beta\bigr).
\end{eqnarray}
Each of these energies gives the same growth rate $\tau$ from
Eq.~\eqref{e:taudef} for solutions that grow exponentially.  At present we
have not found a use for this unexpected abundance of symmetrizers and
energy norms.  In our numerical work below we choose (fairly
arbitrarily) one member of this family to compute our growth-rate
estimates.

In our earlier discussion we found it useful to analyze the
symmetrizers associated with these equations in two steps: first, to
consider the symmetrizers associated with the fundamental
representations of the theory, and second to work out how the general
symmetrizer can be obtained from the fundamental representation by
performing a suitable transformation.  Here we find it useful to
consider the corresponding questions for the energies.  First we
recall that the dynamical fields $u^\alpha$ are related to the
fundamental fields by the transformation,
\begin{equation}
u^\alpha=T^\alpha{}_\beta u^\beta_0,
\end{equation}
where the matrix $T^\alpha{}_\beta$ [defined in Eqs.~\eqref{e:pdef}
and \eqref{e:mdef}] depends on the kinematical parameters and the
metric $g_{ij}$.   Thus perturbations $\delta u^\alpha$ are related to
perturbations in the fundamental fields by:
\begin{eqnarray}
\delta u^\alpha &=& \left(T^\alpha{}_\beta+
\partial_\beta T^\alpha{}_\gamma u^\gamma_0\right) 
\delta u^\beta_0,\nonumber\\
&=& T^\alpha{}_\mu\left(\delta^\mu{}_\beta
+V^\mu{}_\beta\right)\delta u^\beta_0,
\end{eqnarray}
where 
\begin{equation}
V^\alpha{}_\beta=T^{-1\alpha}{}_\gamma \partial_\beta
T^\gamma{}_\sigma u^\sigma_0.\label{e:Ddef}
\end{equation}

One can now work out the transformation properties of the
energy and flux using Eqs.~\eqref{e:atransform} and \eqref{e:stransform}:
\begin{eqnarray}
\delta E &=& S_{\alpha\beta}\delta u^\alpha\delta u^\beta,\nonumber\\
&=& \delta E_0 + S^0_{\mu\nu}V^{\mu\nu}_{\alpha\beta}
\delta u^\alpha_0\delta u^\beta_0,\label{e:de0trans}
\end{eqnarray}
where $V^{\mu\nu}_{\alpha\beta}$ is defined as
\begin{equation} V^{\mu\nu}_{\alpha\beta}= \delta ^\mu_{\alpha}
V^{\nu}{}_{\beta}+\delta ^\mu_{\beta}
V^{\nu}{}_{\alpha}+V^{\mu}{}_{\alpha}V^{\nu}{}_{\beta}.
\label{e:DDdef}
\end{equation}
Thus the energy $\delta E$ is just the fundamental energy $\delta E_0
=S^0_{\alpha\beta}\delta u^\alpha_0\delta u^\beta_0$ plus terms
proportional to $V^{\mu\nu}_{\alpha\beta}$.  A similar argument leads to the
transformation for the energy flux:
\begin{equation}
\delta E^n= \delta E^n_0 + A^{n}_0{}_{\mu\nu}
V^{\mu\nu}_{\alpha\beta}\delta u^\alpha_0\delta u^\beta_0.
\label{e:de0ktrans}
\end{equation}
We note that the only dependence of the energy and flux on the
kinematical parameters comes through $V^{\mu\nu}_{\alpha\beta}$ and hence
$V^{\mu}{}_{\alpha}$.

Finally, we note that the expression for the transformation of the
terms $C_{\alpha\beta}\delta u^\alpha\delta u^\beta$ can be obtained
by a similar calculation.  The result is
\begin{eqnarray}
&&\!\!\!\!\!\!\!\!
C_{\alpha\beta}\delta u^\alpha\delta u^\beta=
\biggl[C^0_{\alpha\beta}+
2S^0_{\mu\nu}V^{\mu\nu}_{\alpha\sigma}F^\sigma_{0\beta}
+\partial_t\left(S^0_{\mu\nu}V^{\mu\nu}_{\alpha\beta}
\right)\nonumber\\
&&\qquad\qquad
+(\sqrt{g})^{-1}\partial_n\left(\sqrt{g}A^{n}_{0\mu\nu}V^{\mu\nu}_{\alpha\beta}
\right)+E_{\alpha\beta}\biggr]
\delta u^\alpha_0\delta u^\beta_0,\nonumber\\\label{e:ctrans}
\end{eqnarray}
where the term $E_{\alpha\beta}$ is defined by
\begin{eqnarray}
&&\!\!\!\!
E_{\alpha\beta}\delta u^\alpha_0\delta u^\beta_0
=
-2\left(\delta^\mu_\alpha+V^\mu{}_\alpha\right)\delta u^\alpha_0
S^0_{\mu\nu}V^\nu{}_\sigma A^{n\sigma}_{0}{}_\beta
\partial_n\delta u^\beta_0
\nonumber\\
&&\qquad\quad\,
+2\left(\delta^\mu_\alpha+V^\mu{}_\alpha\right)\delta u^\alpha_0
A^n_{0\mu\nu}V^\nu{}_\beta\partial_n\delta u^\beta_0.\label{e:eab}
\end{eqnarray}
This expression is obtained by straightforward calculation using
Eqs.~\eqref{e:de0trans} through \eqref{e:de0ktrans}.  Note that the
left side of Eq.~\eqref{e:eab} is a quadratic form in $\delta
u^\alpha_0$ while the right side depends on the derivatives
$\partial_n\delta u^\alpha_0$.  To understand this we use the fact
that $V^\mu{}_\alpha$ is non-zero only when the index $\alpha$
corresponds to one of the spatial metric components,
i.e. $V^\alpha{}_\beta \delta u ^\beta_0=V ^{\alpha ij}\delta g
_{ij}$.  This follows from Eqs.~\eqref{e:vaij} and \eqref{e:Ddef} and
the fact that $T^\alpha{}_\beta$ depends on $g_{ij}$ but none of the
other dynamical fields.  It follows that the term $V^\nu{}_\sigma
A^{n\sigma}_0{}_\beta\partial_n\delta u^\beta_0$ includes only the
derivatives of fields that are present in the metric evolution
equation.  But this evolution equation, Eq.~\eqref{ea:kstg}, includes
only the derivatives of the metric along the shift vector, and so it
follows that $V ^\nu{}_\sigma A ^{k\sigma}_0{}_\beta= -N ^k V
^\nu{}_\beta $.  Thus the expression for $E_{\alpha\beta}$ may be
written in the form:
\begin{eqnarray}
E_{\alpha\beta}\delta u^\alpha_0\delta u^\beta_0
&=&4\left(\delta^\mu_\alpha+V ^\mu{}_\alpha\right)
\left(A ^k_{0\mu\nu}+N ^kS^0_{\mu\nu}\right)\delta u ^\alpha_0\nonumber\\
&&\times
V ^{\nu ij}\delta D _{kij}
\label{e:edef}
\end{eqnarray}
The last terms in this expression came from the term $V^\nu{}_\beta
\partial _k\delta u^\beta_0$, which [using Eq.~\eqref{e:vaij}] depends
only on the spatial derivatives of the metric perturbations $\delta
g_{ij}$: $V ^\nu{}_\beta \partial _k\delta u^\beta_0= V^{\nu
ij}\partial _k\delta g _{ij}=2V^{\nu ij}\delta D _{kij}$.  Thus the
right side of Eq.~\eqref{e:edef} is a quadratic form in the $\delta
u^\alpha_0$ as required. Finally we note that $C_{\alpha\beta}\delta
u^\alpha\delta u^\beta$, like the energy and energy-flux, depends on
the kinematical parameters only through $V^\alpha{}_\beta$.  The
expression for the transformation of the matrix $\bar C
_{\alpha\beta}$ that appears in our approximate expressions for the
instability growth-rates, Eq.~\eqref{e:tauapproxdef2}, follows
directly from Eq.~\eqref{e:ctrans}:
\begin{eqnarray}
&&\!\!\!\!\!\!\!\!  \bar C _{\alpha\beta}\delta u^\alpha\delta
u^\beta= \biggl[\bar C ^0_{\alpha\beta}+
2S^0_{\mu\nu}V^{\mu\nu}_{\alpha\sigma}F^\sigma_{0\beta}
+\partial_t\left(S^0_{\mu\nu}V^{\mu\nu}_{\alpha\beta}
\right)\nonumber\\ &&\qquad\qquad\qquad +E_{\alpha\beta}
- 2 k_n A^{n}_{0\mu\nu}V^{\mu\nu}_{\alpha\beta}
\biggr] \delta
u^\alpha_0\delta u^\beta_0.
\end{eqnarray}
%

\section{Numerical Tests}
\label{s:growthrates}

In this section we compare the approximate expressions for the
instability growth rates developed in Sec.~\ref{s:energynorms} with
growth rates determined directly from numerical solutions of the
Einstein evolution equations.  For this study we use the 2-parameter
subset of the KST equations, discussed in Sec.~\ref{s:symmetric},
referred to as the generalized Einstein-Christoffel
system~\cite{Kidder2001}.  Since we do not yet understand the meaning
of the different energy norms developed in
Sec.~\ref{s:kstenergynorms}, we limit our consideration here to the
norm computed from the symmetrizer with $A_1=B^0_1=1/3$ and
$A_2=B^0_2=1$.  This choice is the closest analog we have of the
simple ``Euclidean'' metric of Eq.~\eqref{e:euclid} for these systems.

We examine the accuracy of the approximate expression for the growth
rate, Eq.~\eqref{e:tauestimate}, by examining the evolution of
perturbations about two rather different background spacetimes: flat
spacetime in Rindler coordinates~\cite{Wald}, and the Schwarzschild
geometry in Painlev\'e-Gullstrand
coordinates~\cite{painleve21,gullstrand22,Martel2000}.  In each of
these cases the full 3D numerical evolution of these equations has
constraint-violating (and possibly gauge) instabilities, and the
approximate expressions for the growth rates using our new formalism
are simple enough that we can evaluate them in a straightforward
manner (even analytically in some cases).  Thus we are able to compare
the estimates of the growth rates with the full numerical evolutions
in a systematic way.

\subsection{Rindler Spacetime}
\label{sec:rindler-spacetime}

\begin{figure}
\begin{center}
\includegraphics[width=2.5in]{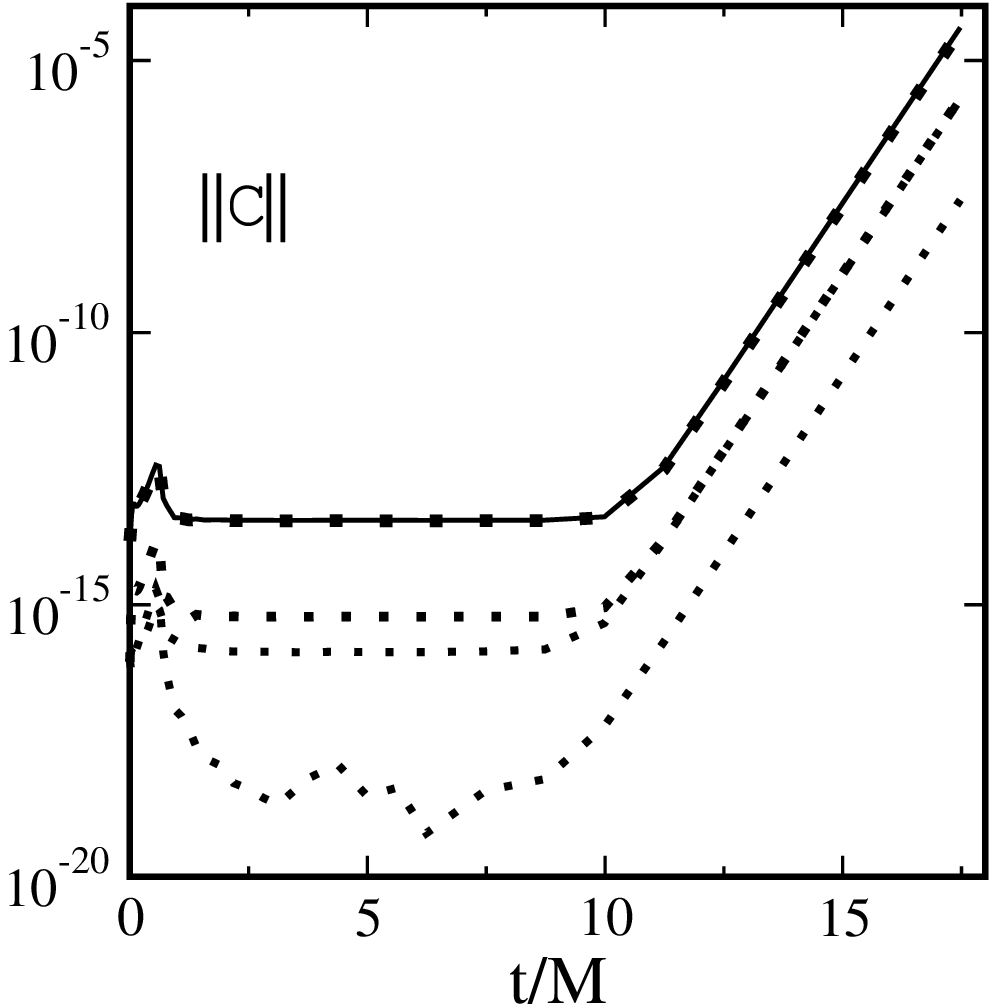}
\end{center}
\caption{Solid curve shows the evolution of the sum of the integral norms of 
all of the constraints.  Dotted curves show the individual
contributions from the various constraints: ${\cal C}
_{klij}{\cal C} ^{klij}$, ${\cal C}{}^2$,
${\cal C} _{kij}{\cal C} ^{kij}$, and ${\cal C}
_{k}{\cal C} ^{k}$ (in that order from largest to
smallest).  }
\label{f:fig1a}
\end{figure}

Flat spacetime can be expressed in Rindler coordinates as follows,
\begin{equation}
ds^2=-x^2 dt^2 + dx^2 +dy^2 +dz^2.
\end{equation}
In this geometry the dynamical fields have the following simple forms,
\begin{eqnarray}
\label{eq:rindler3plus1start}
g_{ij}&=&\delta_{ij},\\
K_{ij}&=&0,\\
D _{kij}&=&0,\\
N&=&x,\\
\label{eq:rindler3plus1end}
N ^k&=&0,
\end{eqnarray}
where $\delta_{ij}$ represents the Euclidean metric in Cartesian
coordinates.  Figure~\ref{f:fig1a} illustrates that even this simple
representation of flat space is subject to the constraint violating
instabilities.  This figure shows the evolution of the
norms of each of the constraints defined in Eqs.~\eqref{e:ch} through
\eqref{e:ct}.  This figure also illustrates that all of the constraints
grow at the same exponential rate in our simulations.

The simplicity of the expressions~\eqref{eq:rindler3plus1start}
through \eqref{eq:rindler3plus1end} allows us to evaluate the various
quantities needed to make our stability estimates in a reasonably
straightforward way.  The first consequence of this simple form is
that the unperturbed fundamental fields consist only of metric fields:
$u^\alpha_0=\{\delta_{ij},0,0\}$.  This fact considerably simplifies
many of the needed expressions.  In particular, the quantity $\partial
_\alpha T ^\mu{}_\nu u ^\nu_0$ vanishes identically for this geometry
because $T^\mu{}_\nu$, when $\nu$ has values that correspond to the
components of the metric, is just the identity transformation and
hence independent of the dynamical fields.  Thus the matrix
$V^\mu{}_\alpha$ defined in Eq.~\eqref{e:Ddef} vanishes identically
for the Rindler geometry.  It follows that for Rindler $\delta
E=\delta E_0$, $\delta E^n=\delta E^n_0$, and $ C_{\alpha\beta}\delta
u^\alpha\delta u^\beta= C^0_{\alpha\beta}\delta u^\alpha_0\delta
u^\beta_0$.  Thus the timescale $1/\tau$ associated with the
instability in the Rindler geometry is completely independent of the
kinematical parameters of the representation of the evolution
equations.  This independence of $1/\tau$ on the kinematical
parameters has been verified numerically for the 2-parameter
generalized Einstein-Christoffel subset of the linearized KST
equations.

Next we evaluate the simple estimate of the growth rate of the
fastest-growing mode using Eq.~\eqref{e:tauapproxdef}.  First we
evaluate the quadratic form $\bar C^0_{\alpha\beta}\delta
u^\alpha_0\delta u^\beta_0$.  For the Rindler geometry with $k_n=0$
this quantity has a reasonably simple form:
\begin{eqnarray}
\!\!\!\!\!
\!\!\!\!\!
&&\bar C^0_{\alpha\beta}\delta u^\alpha \delta u^\beta=\nonumber\\
&&\quad 2\delta K^{ij}\left[(\zeta+3) \hat x^k\delta D _{ikj}
-2\hat x^k\delta D _{kij}-2x\delta g_{ij}
\right]\quad
\nonumber\\
&&\quad+\hat x^k \delta K^i{}_{k}\bigl\{(Q_1-2)\delta D _{ij}{}^{j}
-[Q_1+2(\zeta-\gamma)]\delta D _{ji}{}^{j}\bigr\}
\nonumber\\
&&\quad
+\hat x^k \delta K^i{}_{i}\bigl[(Q_1+2)\delta D _{jk}{}^j
-Q_2\delta D _{kj}{}^{j}\bigr],\quad
\end{eqnarray}
where $\hat x_i=\nabla_i x$, and $Q_1$ and $Q_2$ are given by
\begin{eqnarray}
Q_1&=&2\gamma(5+3\gamma)+10\frac{1-\zeta}{3\zeta+5},\\
Q_2&=&6\gamma(2+\gamma)+8\frac{\zeta+5}{3\zeta+5}.
\end{eqnarray}
This symmetric quadratic form is simple enough that its eigenvalues
and eigenvectors can be determined analytically~\footnote{We do not
reproduce here the rather complicated expressions for the eigenvalues
of $\bar C _{\alpha\beta}$ in the Rindler geometry, but we would be
happy to supply them to anyone who has an interest in them.}.  All of
these eigenvalues depend on the two dynamical parameters
$\{\zeta,\gamma\}$ but not on any of the kinematical parameters.  The
maximum eigenvalue also depends on position in Rindler space,
according to ${\bar \lambda}_{\max}=f(\zeta,\gamma)+4x^2$, where $x$ is the
Rindler coordinate.  This eigenvalue has the (approximate) minimum
value ${\bar \lambda}_{\max}^2=10.198+4x^2$ at the point
$\{\zeta,\gamma\}=\{-0.135,-1.382\}$.  In our growth rate estimates we
evaluate the eigenvalue at the point in our computational domain where
${\bar \lambda}_{\max}$ has its maximum value, which in our case is at $x=1$.

\begin{figure}
\begin{center}
\includegraphics[width=2.5in]{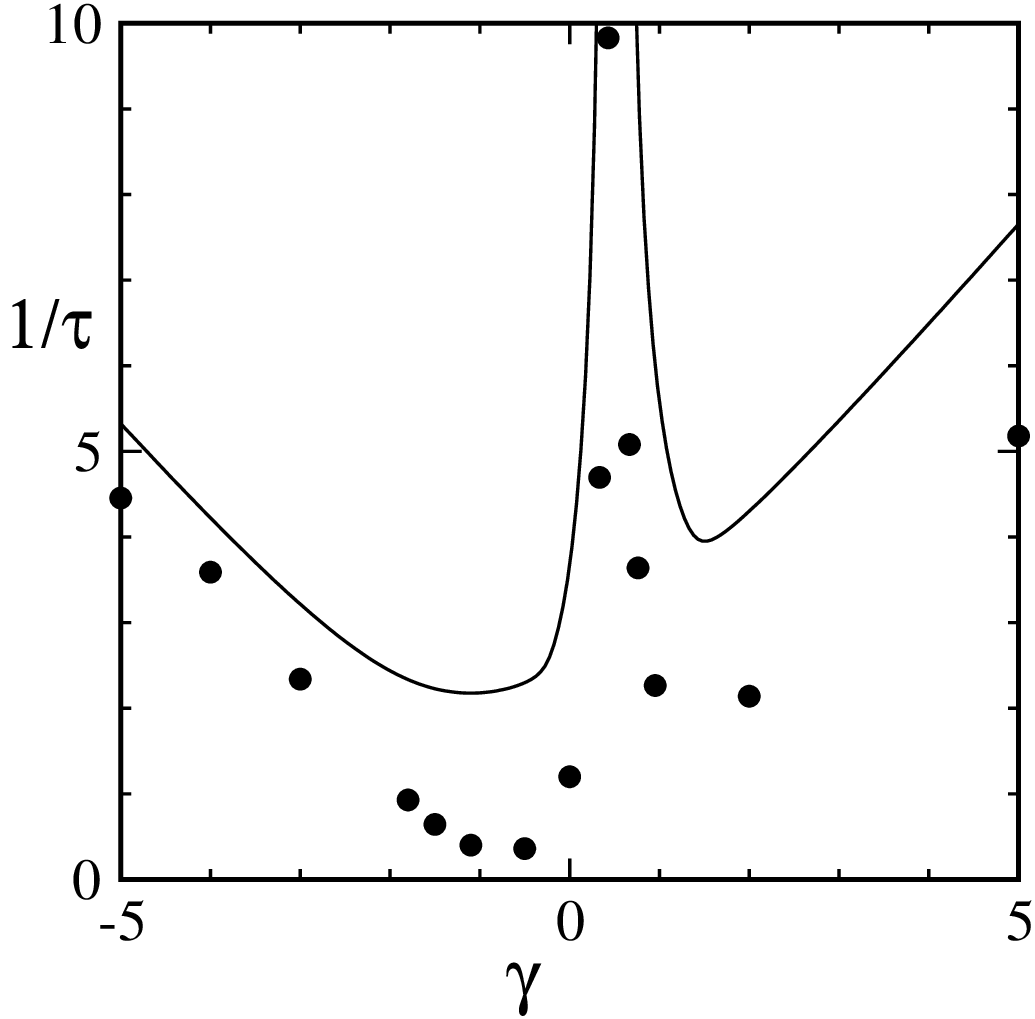}
\end{center}
\caption{Solid curve shows (half) the largest eigenvalue of $\bar C
_{\alpha\beta}$ for the Rindler spacetime (with $k_n=0$).  Dots give
values for the actual growth rate of the short timescale instability
as determined by numerical solution of the linearized evolution
equations with $\zeta=-1$.}
\label{f:fig2}
\end{figure}

\begin{figure}
\begin{center}
\includegraphics[width=2.5in]{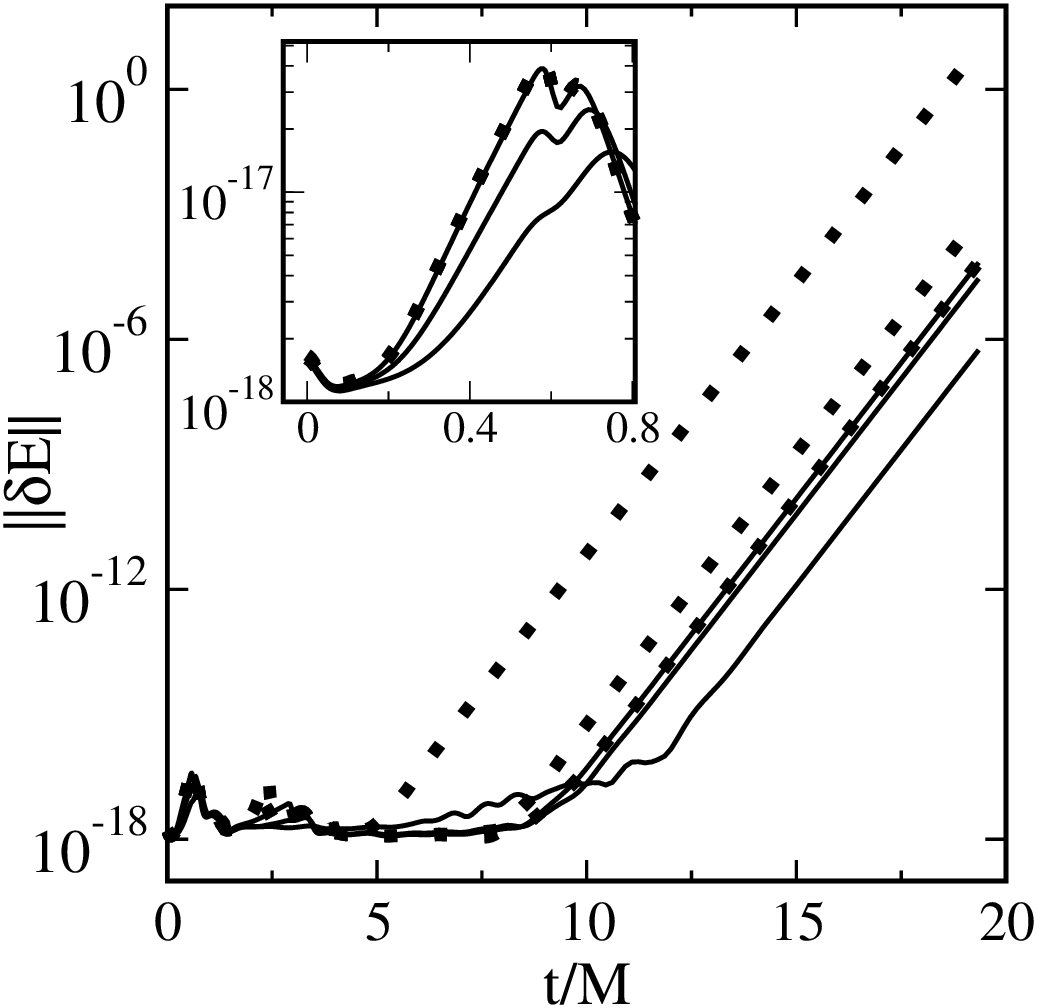}
\end{center}
\caption{Evolution of the energy norm for perturbations of Rindler
space.  The solid curves show the evolutions based on a three
different resolutions (101, 401 and 6401 grid points) of a finite
difference version of the code, while dotted curves show the same
evolution using three different spectral resolutions (64, 96, and 128
basis functions).  A magnified view of the first part of the evolution
(upper left) illustrates the short timescale instability whose growth
rate is shown in Fig.~\ref{f:fig2}.}
\label{f:fig2a}
\end{figure}

We illustrate in Fig.~\ref{f:fig2} the dependence of $\half{\bar
\lambda}_{\max}$ on the parameter $\gamma$ (for $\zeta=-1$).  Our
simple analysis predicts that the best-behaved form of the evolution
equations should be obtained for parameter values near this minimum.
We also plot in Fig.~\ref{f:fig2} numerically-determined points that
represent the growth rates of the instability.  These points were
evaluated from the growth rate of the energy norm for evolutions of
the linearized KST equations using a 1D pseudospectral method on the
domain $x \in [0.1,1]$.  The numerical method (but not the system of
equations) is identical to the one described in \cite{Kidder2000a},
except here our spatial coordinate is the Cartesian coordinate $x$
rather than the spherical coordinate $r$.  All evolutions are
performed at multiple resolutions (see Fig.~\ref{f:fig2a}) to test
convergence, and the growth rates that we quote in the text and in the
figures are taken from the converged solution. At the boundaries we
demand that incoming characteristic fields have zero time derivative,
and we do not constrain the outgoing and nonpropagating characteristic
fields. The initial data is taken to be the exact solution plus a
perturbation of the form $A e^{-(x-0.55)^2/w^2}$ that is added to each
of the dynamical variables ($g_{ij}$,$K_{ij}$,$M_{kij}$).  The
perturbation amplitude $A$ for each variable is a randomly chosen
number in the interval $(-10^{-8},10^{-8})$.  Note that this
perturbation violates the constraints.  The gauge fields $\beta^i$ and
$Q$ are not perturbed.  We measure the growth rates of the energy norm
during the very earliest parts of the evolutions, before the initial
Gaussian pulse can propagate to the edge of the computational
domain. To do this we use an extremely narrow pulse $w=0.0125$, so
that there is sufficient time to measure the growth rate before even
the tail of the pulse reaches the boundary (one could also widen the
computational domain, but this is equivalent to changing the pulse
width because the Rindler solution is scale-invariant).

Figure~\ref{f:fig2} illustrates that the analytical estimate
consisting of $1/\tau\approx \half{\bar \lambda}_{\max}$ gives a
reasonably good estimate of this initial growth rate of the
instability.  And the location of the optimal value of the parameter
$\gamma$, where the growth rate of the instability is minimum, also
agrees fairly well with the location of the minimum of
$\half{\bar \lambda}_{\max}$.

However, the short timescale instabilities illustrated in
Fig.~\ref{f:fig2} are not our primary concern here.
Figure~\ref{f:fig2a} illustrates the evolution of the energy norm for
one of the evolutions discussed above (with $\gamma=\frac{1}{3}$,
$\zeta=-1$, $w=0.025$).  This shows that once the unstable initial
pulse crosses and leaves the computational domain (in about half a
light crossing time, or $t\sim 0.6$), the solution grows much less
rapidly.  While the short term instability does not seriously effect
the long term stability of the code (unless it is large enough that
the code blows up in less than a crossing time), Fig.~\ref{f:fig2a}
illustrates that there are often other instabilities that grow more
slowly, but which can contaminate and eventually destroy any attempts
to integrate the equations for long periods of time.
Figure~\ref{f:fig2a} also illustrates the equivalence between evolving
the system using a 1+1 dimensional finite difference code and a
pseudo-spectral code.  The finite difference code uses a first-order
upwind method (see, {\it e.g.\,} \cite{leveque}) in which the
fundamental variables are decomposed into characteristic fields.  Both
the finite difference and the pseudo-spectral methods use the same
initial data, boundary conditions, and gauge conditions. Three
different spatial resolutions are illustrated for each method, and the
highest resolution runs essentially coincide.  This agreement provides
additional evidence that the instabilities discussed here are features
of the analytical evolution equations, and are not numerical in
origin.

In our estimates, we will now attempt to filter out the less
interesting short term instabilities by imposing boundary conditions
on the trial eigenfunctions $\delta u^\alpha$ used in
Eq.~\eqref{e:tauapproxdef}.  These boundary conditions are implemented
using the projection operator $P ^\mu{}_\nu$ in
Eq.~\eqref{e:tauapproxdef2}.  For the case of the Rindler geometry
this projection operator is constructed to annihilate both the ingoing
and outgoing propagating modes.  Thus the growth rate estimate given
in Eq.~\eqref{e:tauestimate} is implemented by finding the largest
eigenvalue of $\bar C _{\alpha\beta}P ^\alpha{}_\mu P ^\beta{}_\nu$
that is projected onto the subspace of non-propagating modes (i.e. the
eigenvectors of $\hat x_k A ^{k\alpha}{}_\beta$ having eigenvalue
zero, as measured by a hypersurface orthogonal observer).  We find
that the largest eigenvalue of this $\bar C _{\alpha\beta}P
^\alpha{}_\mu P ^\beta{}_\nu$ is zero for all values of
$\{\zeta,\gamma\}$, for all values of the wavevector $k_n$, and for
all values of $x$: the non-propagating modes of Rindler are all stable
according to this estimate.

\begin{figure}
\begin{center}
\includegraphics[width=2.5in]{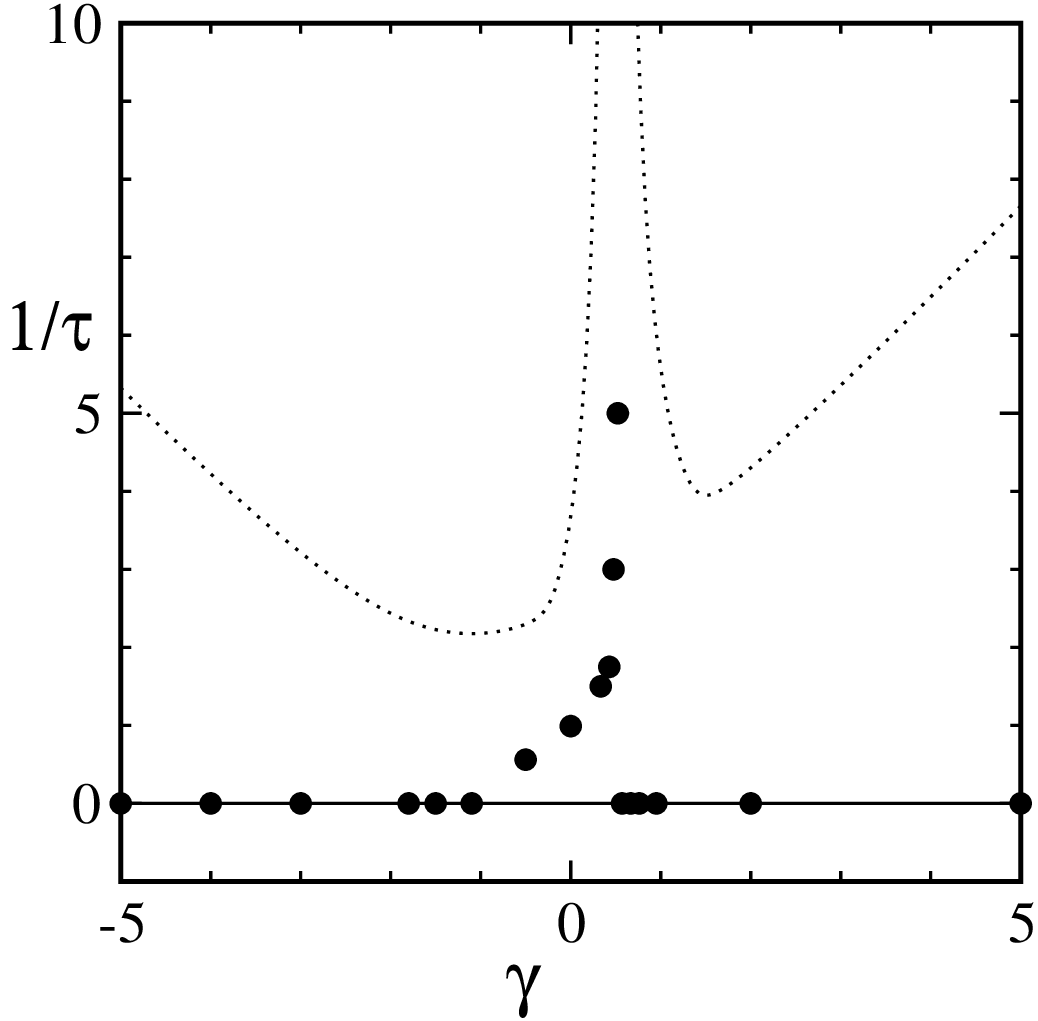}
\end{center}
\caption{Solid curve illustrates the simple analytical growth rate
estimate $1/\tau=0$ for the Rindler spacetime.  Points are
growth rates determined numerically for long-term (many light crossing
times) evolutions with $\zeta=-1$.  Dotted curve represents the
short-term instability growth rate estimate.}
\label{f:fig3}
\end{figure}

This estimate of the long-term growth rate in Rindler is shown as the
solid curve in Fig.~\ref{f:fig3}.  The points in Fig.~\ref{f:fig3}
represent growth rates determined numerically over many many light
crossing times (for these evolutions, we have no need for an extremely
narrow pulse; we use $w=0.1$ so that we can run at lower resolutions).
We observed that these equations are in fact stable for most values of
the parameter $\gamma$ over these timescales.  The only long term
instabilities that we observed in Rindler occur for values of $\gamma$
near 0.5, where the short-term instability growth rate is infinite.
We believe these instabilities occur because of coupling in the
evolution equations between the pure propagating and non-propagating
modes used in our simple estimate.

\subsection{Schwarzschild Spacetime}
\label{sec:schw-spac}

We have also studied instabilities in the
Einstein evolution equations for solutions that are close to the
Schwarzschild geometry.  We use the
Painlev\'e-Gullstrand~\cite{painleve21,gullstrand22,Martel2000} form of the
Schwarzschild metric:
\begin{equation}
ds^2 = -dt^2 + \left(dr+\sqrt{2M/r}dt\right)^2 + r^2 d\Omega^2,
\end{equation}
where $d\Omega^2$ represents the standard metric on the unit sphere.
The dynamical fields that represent this geometry are also quite
simple (although not quite as simple as Rindler).  In Cartesian
coordinates we have
\begin{eqnarray}
\label{eq:paingull1}
g _{ij}&=& \delta _{ij},\\
K _{ij}&=& \sqrt{2M/r^3}\delta _{ij} - 3\sqrt{M/2r^3}\hat r_i \hat
r_j,\\
D _{kij}&=&0,\\
N&=&1,\\
\label{eq:paingull2}
N ^k&=&\sqrt{2M/r}\hat r^k,
\end{eqnarray}
where $\delta _{ij}$ is the Euclidean metric (in Cartesian
coordinates), and $\hat r ^k=(x,y,z)/r$ is the unit vector in the
radial direction.  In this representation of the Schwarzschild
geometry the fundamental representation of the dynamical fields $u
^\alpha_0=\{\delta _{ij},K _{ij},0\}$ includes a non-vanishing
extrinsic curvature.  Therefore $\partial _\mu T ^{\alpha}{}_\beta u
^\beta_0$ will be non-zero for this geometry.  Since the $D _{kij}$
components of $u^\alpha_0$ are still zero, it follows that only the $K
_{ij}$ components of $T ^\alpha{}_\beta$ will contribute to $V
^\mu{}_\alpha$.   One can show that only the $K
_{ij}$ components of $V ^\mu{}_\alpha\delta u^\alpha_0$ are non-zero,
and that these are given by:
\begin{equation}
\frac{\zhat}{1+3\zhat}
\left[(1+3\zhat)K \delta ^a_i \delta ^b_j-\zhat K g _{ij}g ^{ab}
- g _{ij}K ^{ab}\right]\delta g _{ab}.
\end{equation}
Thus $V ^\mu{}_\alpha$ depends only on the kinematical parameter
$\zhat$.  Consequently for evolutions near the representation of the
Schwarzschild geometry considered here, the growth rates of the
instabilities will depend on only three of the
nine KST parameters: $\{\zeta,\gamma,\zhat\}$.

We solve the evolution equations for the perturbed Schwarzschild
geometry using a pseudospectral collocation method (see, {\it e.g.\/}
\cite{Boyd1989,Canuto1988} for a general review,
and~\cite{Kidder2000a,Kidder2001,Kidder2002a} for more details of the
particular implementation used here) on a 3D spherical-shell domain
extending from $r=1.9M$ to $r=11.9M$. Our code utilizes the method of
lines; the time integration is performed using a fourth order
Runge-Kutta algorithm. The inner boundary lies inside the event
horizon; at this boundary all characteristic curves are directed out
of the domain (into the black hole), so no boundary condition is
required there and none is imposed (``horizon excision''%
~\cite{seidel_suen92,alcubierre94,%
anninos_etal95,Gundlach1999,scheel_etal97a,scheel95a,scheel95b,%
bbhprl98a,Brandt2000,Alcubierre2000,Alcubierre2001A,%
Kidder2001}).
This reflects the causality condition that the interior of a black
hole cannot influence the exterior region.  At the outer boundary we
require that all ingoing characteristic fields be time-independent,
but we allow all outgoing characteristic fields to propagate freely.
The initial data is the exact solution Eqs.~\eqref{eq:paingull1}
through \eqref{eq:paingull2}, plus a perturbation of the form $A
e^{-(r-7M)^2/4M^2}$ added to each of the 30 dynamical variables (the
Cartesian components of $g_{ij}$, $K_{ij}$, and $M_{kij}$). The
perturbation amplitude $A$ for each variable is a randomly chosen
number in the interval $(-10^{-8},10^{-8})$.  The gauge fields
$\beta^i$ and $Q$ are not perturbed.  Because we perturb the {\it
Cartesian\/} components of each field by a spherically symmetric
function, the initial data are not spherically symmetric.  Note that
we have chosen an initial perturbation that violates the constraints.

\begin{figure}
\begin{center}
\includegraphics[width=2.5in]{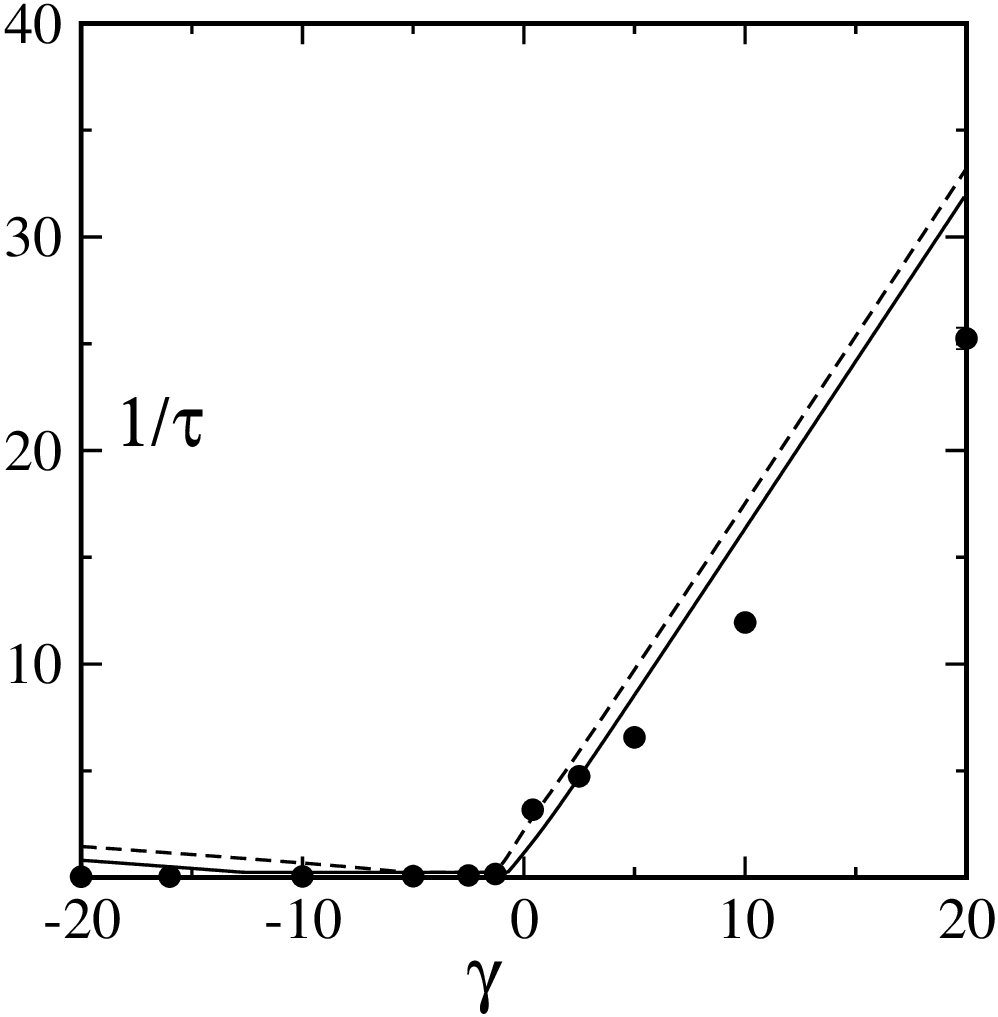}
\end{center}
\caption{Dashed curve illustrates the simple analytical growth rate
estimate for the Schwarzschild spacetime with $k_n=0$, and the solid
curve shows estimate for $k_n=-\hat r_n/M$.  Points are growth rates
determined numerically for long-term (many light crossing times)
evolutions of the 3D linearized equations with $\zeta=-1$ and
$\zhat=-1/4$.}
\label{f:fig4}
\end{figure}

\begin{figure}
\begin{center}
\includegraphics[width=2.5in]{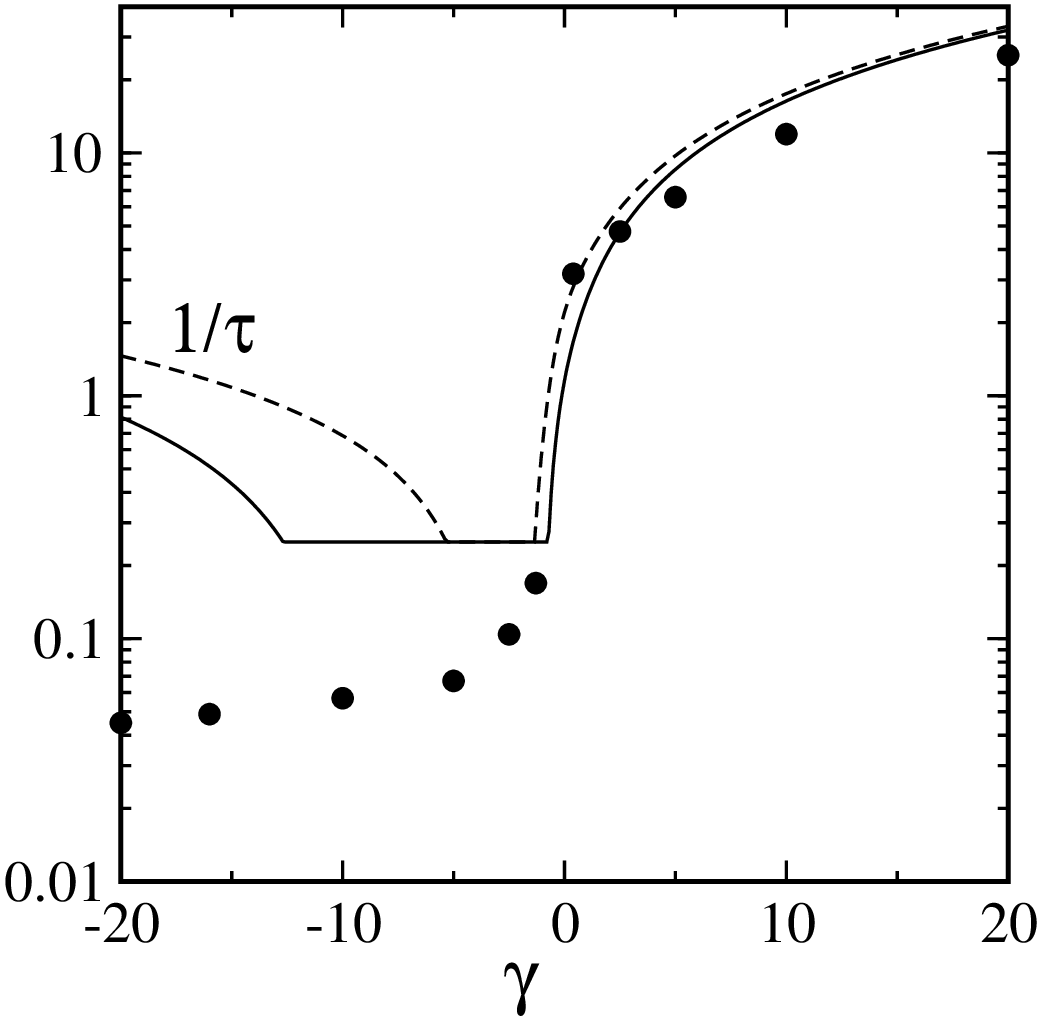}
\end{center}
\caption{Same information as Fig.~\ref{f:fig4} but plotted here using
a logarithmic scale.  This illustrates that while the simple
analytical growth rate estimate is qualitatively correct, it fails to
correctly identify the optimum choice of parameters.}
\label{f:fig5}
\end{figure}

At the outer boundary, the modes that appear non-propagating to a
hypersurface orthogonal observer actually are incoming with respect to
the boundary because of the outward-directed radial shift vector
$N^i$. Thus the projection operator $P ^\mu{}_\nu$ needed to construct
our growth rate estimates from Eq.~\eqref{e:tauestimate} is therefore
the one that annihilates both the incoming and the ``non-propagating''
modes, but leaves the outgoing modes unchanged.  We have computed the
eigenvalues of this appropriately projected $\bar C _{\alpha\beta}$,
and illustrate the largest eigenvalue in Fig.~\ref{f:fig4}.  These
eigenvalues depend on the radial coordinate $r$ in this spacetime, and
we plot in Fig.~\ref{f:fig4} the largest value of this eigenvalue,
which occurs at $r=2M$.   The graph represents (half) this largest
eigenvalue as a function of the parameter $\gamma$ for $\zeta=-1$ and
$\zhat=-1/4$.  In Fig.~\ref{f:fig4} we give estimates for two choices
of the wavevector $k_a$ that characterizes the spatial variation of
$\delta u^\alpha$: the dashed curve represents the choice $k_n=0$,
while the solid curve represents the choice $k_n=-\hat r_n/M$.  The
points in this graph represent the numerically-determined growth rates
of the instabilities for the linearized equations.  We see that the
agreement with our very simple analytical estimate is again quite
good.  However, Fig.~\ref{f:fig5} illustrates that while the simple
analytical estimate is correct qualitatively, it fails to correctly
identify the best choice of parameters to use for long-term numerical
evolutions.  The minimum growth rate is actually smaller (by about a
factor of two) than what could be achieved by the optimal range of
parameters that are identified by the simple analytical growth rate
estimate.

\begin{figure}
\begin{center}
\includegraphics[width=2.5in]{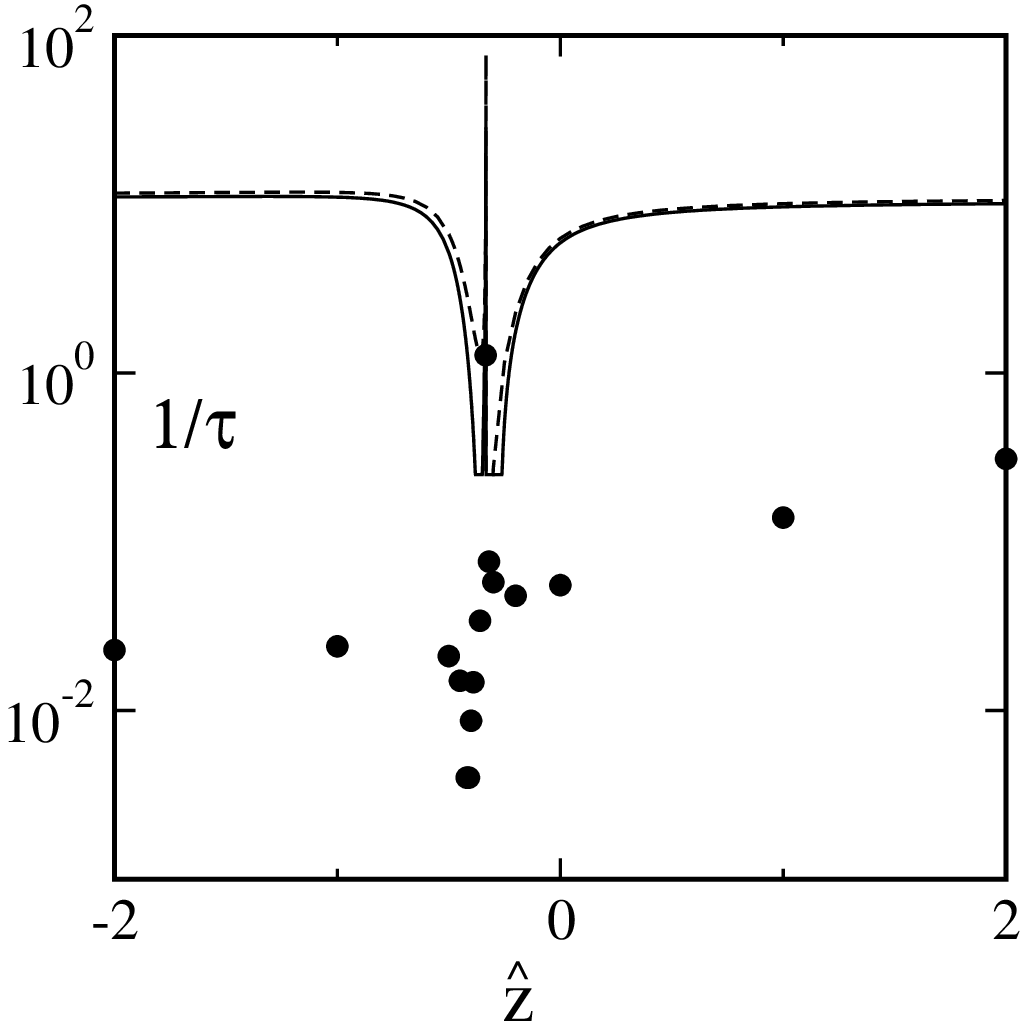}
\end{center}
\caption{Dashed curve illustrates the simple analytical growth rate
estimate for the Schwarzschild spacetime with $k_n=0$, and the solid
curve shows the estimate for $k_n=-\hat r_n/M$.  Points are growth
rates determined numerically for long-term (many light crossing times)
3D evolutions with $\zeta=-1$ and $\gamma=-16$.}
\label{f:fig6}
\end{figure}

Fig.~\ref{f:fig6} is the same as Fig.~\ref{f:fig5} except the results
are plotted as a function of the parameter $\zhat$ for $\zeta=-1$ and
$\gamma=-16$. Again, the actual minimum growth rate is smaller than
the analytical estimate. However, the qualitative shape of the
analytical curve is correct, and can be used as a guide for choosing
parameters to investigate with the numerical code. In the present case
this guide proves extremely useful, for it has allowed us to find a
parameter choice ($\zhat=-0.42$, $\gamma=-16$, $\zeta=-1$) that
significantly extends the amount of time the fully 3D nonlinear code
can evolve a single Schwarzschild black hole.  The very narrow range
of parameters where the evolution equations have optimal stability, as
shown in Fig.~\ref{f:fig6}, illustrate why these optimal values were
not discovered empirically~\cite{Kidder2001}.  The growth rate
estimates derived here were needed to focus the search onto the
relevant part of the parameter space.  The improvement in 3D nonlinear
black hole evolutions resulting from these better parameter choices
will be discussed in detail in a forthcoming paper.


\section{Discussion}
\label{s:discussion}

This paper studies the constraint-violating and gauge instabilities of
the Einstein evolution equations.  We derive an analytical expression
for the growth rate of an energy norm in
Sec.~\ref{s:energynorms}. This energy norm measures the deviations of
a given solution from a background constraint-satisfying solution.  We
show numerically that the growth rate of this energy norm is identical
to the growth rate of constraint violations in solutions of both the
linearized equations and of the full nonlinear equations.  Thus we
concentrate here on the analysis of the evolution of this energy norm.
Section~\ref{s:kstequations} derives the analytical expressions needed
to evaluate this energy norm in the 12-parameter family of Einstein
evolution equations introduced by Kidder, Scheel, and
Teukolsky~\cite{Kidder2001}.  This analysis demonstrates that an open
subset of the KST equations with all physical characteristic speeds is
symmetric hyperbolic.  And perhaps more surprisingly, there is a large
open subset of these strongly hyperbolic equations that is not
symmetric hyperbolic.

We make a considerable effort in this paper to demonstrate that the
instabilities that we observe are actual solutions to the evolution
equations and not numerical artifacts of the discrete representation
of the solution that we use.  We run many of the cases that we study
here using both a finite difference and a pseudo-spectral version of
the code, and find equivalent results.  We run all numerical
simulations at multiple spatial and temporal resolutions to
demonstrate that numerical convergence to the desired accuracy is
achieved.  We also confirm that the growth rate determined from our
analytical integral expression is equal (to the desired numerical
accuracy) to the growth rate measured numerically from the growth of
the energy norm.  This equality is expected only for solutions to the
actual evolution equations, so this provides a non-trivial check that
the instabilities we find are real solutions to the equations.

The analysis presented here also suggests that the instabilities that
we see are endemic to the Einstein evolution equations and are not the
result of improper boundary conditions.  We impose conditions that
suppress the incoming components of the dynamical fields at the
boundaries of the computational domain.  These conditions (sometimes
called maximally dissipative~\cite{Szilagyi2002}) ensure that the
energy norm does not grow due to energy being inserted into the
computational domain across the boundaries.  Any other boundary
conditions that might be imposed (including ones that attempt to
control the influx of constraint violations~\cite{Calabrese2001})
could only increase the growth rate of our energy norm.  Because the
constraint violations grow at the same rate as the energy norm, it
seems very unlikely that the constraint violations could be eliminated
or even significantly reduced simply by changing the boundary
conditions.  And even if the constraint violations could be eliminated
with better boundary conditions, our analysis shows that
something---presumably gauge instabilities---would still cause the
energy norms to grow on the timescales illustrated here.

The analysis presented here provides additional support for the claim
that the growth rate of instabilities is strongly affected by the
choice of representation of the Einstein evolution equations.  We find
that this growth rate varies considerably as the parameters in the KST
formulation of the equations are varied.  Further, we demonstrate here
that the functional dependence of the growth rate on the parameters
strongly depends on the initial data that are being evolved.  We show
that the functional dependence on the set of parameters is not the
same for the Schwarzschild geometry as it is for flat spacetime in
Rindler coordinates. This result strongly suggests that analyzing the
stability of the evolution equations for perturbations of flat
spacetime in Minkowski coordinates~\cite{Alcubierre2000a,Yoneda2002b},
although useful for screening out particularly poorly-behaved
formulations, is unlikely to succeed in identifying the best form of
the equations to use for binary black hole spacetime evolutions.
Rather these results suggest that it may be necessary to choose the
optimal form of the evolution equations individually for each problem.
Estimates of the instability growth rates such as those found here
(and hopefully more refined estimates yet to be discovered) depend
only on the fields at a given instant of time.  It might be possible
(or even necessary) then to use such estimates to determine and then
adjust the form of the evolution equations to be optimal as the
spacetime evolves.

\acknowledgments We thank H. Friedrich, J. Isenberg, M. Keel,
L. Kidder, L. Lehner, V. Moncrief, R. O'Shaughnessy, H. Pfeiffer,
B. Szilagyi, S. Teukolsky, M. Tiglio and J.  Winicour for helpful
discussions concerning this work.  We also thank L. Kidder,
H. Pfeiffer, and S. Teukolsky for allowing us to use the Cornell code
to test our growth rate estimates.  Some computations were performed
on the IA-32 Linux cluster at NCSA. This research was supported in
part by NSF grant PHY-0099568 and NASA grant NAG5-10707.

\appendix*
\section{}
\label{s:appendix}
In this Appendix we give the details of a number of equations that
define the KST~\cite{Kidder2001} formulation of the Einstein evolution
equations.  The principal parts of these equations
can be written in the form of a first order system:
\begin{equation}
\partial _t u ^\alpha + A ^{k\alpha}{}_\beta\partial _k u ^\beta
\simeq 0.
\end{equation}
Here the dynamical fields are taken to be the set $u ^\alpha=\{g
_{ij}, P _{ij}, M _{kij}\}$, where $g _{ij}$ is the spatial metric,
and $P _{ij}$ and $M _{kij}$ are fields that initially will be
interpreted as the extrinsic curvature $K _{ij}$ and the spatial
derivatives of the metric $D _{kij}=\half\partial _k g _{ij}$
respectively. The matrices $A ^{k\alpha}{}_\beta$ can be written
quite generally for these systems in the form
\begin{eqnarray} \partial{}_t g{}_{ij}&\simeq& N ^n\partial _n g
_{ij},\label{e:kstga}\\ \partial{}_t P {}_{ij}&\simeq& N ^n\partial _n
K _{ij}-N\Bigl[ \mu{}_1 g{}^{nb}\delta{}^c{}_i\delta{}^d{}_j +\mu{}_2
g{}^{nd}\delta{}^b{}_{(i}\delta{}^c{}_{j)} \nonumber\\ &&+\mu{}_3
g{}^{bc}\delta{}^n{}_{(i}\delta{}^d{}_{j)} +\mu{}_4
g{}^{cd}\delta{}^n{}_{(i}\delta{}^b{}_{j)}\nonumber\\ &&+\mu{}_5
g{}^{nd}g{}^{bc}g{}_{ij} +\mu{}_6 g{}^{nb}g{}^{cd}g{}_{ij}\Bigr]
\partial{}_k M {}_{bcd},\label{e:kstpa}\\ \partial{}_t M{}_{kij}
&\simeq& N ^n \partial _n M _{kij} -N\Bigl[ \nu{}_1
\delta{}^n{}_k\delta{}^b{}_i\delta{}^c{}_j
+\nu{}_2\delta{}^n{}_{(i}\delta{}^b{}_{j)}\delta{}^c{}_k\nonumber\\
&&+\nu{}_3 g{}^{nb}g{}_{k(i}\delta{}^c{}_{j)} + \nu{}_4
g{}^{nb}g{}_{ij}\delta{}^c{}_{k}\nonumber\\ &&+ \nu{}_5
g{}^{bc}g{}_{k(i}\delta{}^n{}_{j)} + \nu{}_6
g{}^{bc}g{}_{ij}\delta{}^n{}_{k}\Bigr]\partial{}_n P {}_{bc}.
\label{e:kstma}
\end{eqnarray}
For the fundamental representations of the theory discussed in
Sec.~\ref{s:kstequations} where the dynamical fields are $u
^\alpha_0=\{g _{ij}, K _{ij}, D _{kij}\}$, the constants $\mu_A$ and
$\nu_A$ are related to the 5 dynamical KST parameters
$(\gamma,\zeta,\eta,\chi,\sigma)$ by,
\begin{eqnarray}
\mu_1&=&\nu_1=1,\label{e:mu10}\\
\mu_2&=&-1-\zeta,\\
\mu_3&=&-1+\zeta,\\
\mu_4&=&1+2\sigma,\\
\mu_5&=&-\mu_6=-\gamma,\\
\nu_2&=&0,\\
\nu_3&=&-\nu_5=-\half\eta,\\
\nu_4&=&-\nu_6=-\half\chi
\label{e:nu60}.
\end{eqnarray}

KST~\cite{Kidder2001} generalize this fundamental representation of the
theory by introducing a 7-parameter family of transformations on the
dynamical fields.  These transformations define the new fields $P
_{ij}$ and $M _{kij}$ in terms of the fundamental fields $K _{ij}$ and
$D _{kij}$ via an equation of the form
\begin{equation}
u ^\alpha=T ^\alpha{}_\beta u ^\beta_0.
\end{equation}
The explicit form of this transformation is given in Eqs.~\eqref{e:pdef} and
\eqref{e:mdef}.  Here we give explicit expressions for the inverse
transformation: 
\begin{eqnarray}
&&K_{ij}=P _{ij}+\zbar\, g_{ij}g^{ab} P _{ab},\label{e:pinvdef}\\
&&D _{kij}=\bigl[\kbar\, \delta{}^a_k\delta{}^b_i\delta{}^c_j
+\ebar\, \delta{}^a_{(i}\delta{}^b_{j)}\delta{}^c_k
+\abar\, g{}_{ij} g^{bc}\delta{}^a_k\qquad\qquad\nonumber\\
&&\quad\qquad+\bbar\, g{}_{ij} g^{ab}\delta{}^c_k
+\cbar\, g{}_{k(i} \delta{}^a_{j)}g^{bc}
+\dbar\, g{}_{k(i} \delta{}^c_{j)}g^{ab}
\bigr]M _{abc},\nonumber\\\label{e:minvdef}
\end{eqnarray}
where the constants $\{\zbar,\kbar,\abar,\bbar,\cbar,\dbar,\ebar\}$ are
functions of the kinematical parameters
$\{\zhat,\khat,\ahat,\bhat,\chat,\dhat,\ehat\}$:
\begin{eqnarray}
\zbar&=&-\zhat/(1+3\zhat),\\
\delta \abar & = & 2(3\ehat+4\khat)(\bhat\chat-\ahat\dhat) 
- (\ahat-\bhat-\chat+\dhat)\ehat^2 \qquad\nonumber\\
&&\qquad\qquad-2(2\ahat\ehat-\bhat\ehat-\chat\ehat+2\ahat\khat)\khat,\qquad\\
\delta \bbar& = & 4(2\ehat + \khat)(\ahat\dhat-\bhat\chat)
+2(\ahat-\chat)\ehat^2\nonumber\\
&&\qquad\qquad+2(2\ahat\ehat-\bhat\ehat+\dhat\ehat-2\bhat\khat)\khat,\qquad\\
\delta\cbar & = &  4(2\ehat + \khat)(\ahat\dhat-\bhat\chat)
+2(\ahat-\bhat)\ehat^2\nonumber\\
&&\qquad\qquad+2(2\ahat\ehat-\chat\ehat+\dhat\ehat-2\chat\khat)\khat,\qquad\\
\delta\dbar & = &4(\ehat + 3\khat)(\bhat\chat-\ahat\dhat)
-4\ahat\ehat^2\nonumber\\
&&\qquad\qquad+4(\bhat\ehat+\chat\ehat-\dhat\khat)\khat,\qquad\\
\delta_0\ebar&=&2\ehat,\\
\delta_0\kbar&=&-\ehat-2\khat,\\
\delta_0 & = & \ehat^2 - \ehat\khat - 2\khat^2,\\
\delta & = & \delta_0\bigl[10(\bhat\chat-\ahat\dhat)
+(3\bhat+3\chat-\ahat+\dhat+\ehat)\ehat\nonumber\\
&&\qquad-(6\ahat+2\bhat+2\chat+4\dhat+\ehat+2\khat)\khat\bigr].
\end{eqnarray}
The transformation $T ^\alpha{}_\beta$ is invertible as long as
$\delta\neq0$ and $\zhat\neq -\third$.  In this generic case we may
write
\begin{equation} 
u^\alpha_0=T^{-1\,\alpha}{}_\beta u^\beta,\label{e:uinvtrans}  
\end{equation}
and $T^\alpha{}_\gamma T^{-1\, \gamma}{}_\beta=\delta^\alpha{}_\beta$,
where $\delta^\alpha{}_\beta$ is the identity matrix on the space of
fields.

We have seen in Sec.~\ref{s:kstequations} that the general expression
for the matrices $A^{k\alpha}{}_\beta$ is related to its form in
the fundamental representation of the equations,
$A^{k}_0{}^{\alpha}{}_\beta$, by the simple transformation law:
$A^{k\alpha}{}_\beta=T ^\alpha{}_\mu A^{k}_0{}^{\mu}{}_\nu T
^{-1\,\nu}{}_\beta$.  This transformation may also be expressed somewhat
more concretely as a transformation on the constants $\mu_A$ and
$\nu_A$ that define the matrices $A ^{k\alpha}{}_\beta$.  The
resulting expressions for these constants after the kinematical
transformation are:
\begin{eqnarray}
\mu_1&=&\kbar-\half(1+\zeta)\ebar,\label{e:mu1}\\
\mu_2&=&\half(1-\zeta)\ebar-(1+\zeta)\kbar,\\
\mu_3&=&(1+6\sigma)\bbar-(1-\zeta)\kbar
-\half(1-4\sigma-3\zeta)\dbar\nonumber\\
&&+\half(1+4\sigma+\zeta)\ebar,\\
\mu_4&=&(1+6\sigma)\abar+(1+2\sigma)\kbar
-\half(1-4\sigma-3\zeta)\cbar\nonumber\\
&&-\half(1-\zeta)\ebar,\\
\mu_5&=&(1+2\gamma+4\zhat+6\gamma\zhat)(\bbar-\half\dbar)
+2\sigma\zhat(3\bbar+\dbar+\ebar)\nonumber\\
&&-(\gamma+2\zhat+3\gamma\zhat)(\kbar-\half\ebar)
-\half\zeta\dbar,\\
\mu_6&=&(1+2\gamma+4\zhat+6\gamma\zhat)(\abar-\half\cbar)
+2\sigma\zhat(3\abar+\kbar+\cbar)\nonumber\\
&&+(\gamma+2\zhat+3\gamma\zhat)(\kbar-\half\ebar)-\half\zeta\cbar,\\
\nu_1&=&\khat,\\
\nu_2&=&\ehat,\\
\nu_3&=&(1-\eta-\half\chi)\dhat-\half(\eta+3\chi)\chat
-\fourth(\eta+2\chi)\ehat-\half\eta\khat,\nonumber\\
\\
\nu_4&=&(1-\eta-\half\chi)\bhat-\half(\eta+3\chi)\ahat
-\fourth\eta\ehat-\half\chi\khat,\\
\nu_5&=&\half(2+\eta+3\chi+6\zbar+2\eta\zbar+6\chi\zbar)\chat\nonumber\\
&&+\half(2\eta+\chi+2\zbar+4\eta\zbar+2\chi\zbar)\dhat
+\half(\eta+2\eta\zbar)\khat\nonumber\\
&&+\fourth(\eta+2\chi+4\zbar+2\eta\zbar+4\chi\zbar)\ehat,\\
\nu_6&=&\half(2+\eta+3\chi+6\zbar+2\eta\zbar+6\chi\zbar)\ahat\nonumber\\
&&+\half(2\eta+\chi+2\zbar+4\eta\zbar+2\chi\zbar)\bhat\nonumber\\
&&+\fourth(\eta+2\eta\zbar)\ehat+\half(\chi+2\zbar+2\chi\zbar)\khat.
\end{eqnarray}
These expressions are identical to those in KST~\cite{Kidder2001} except for
$\mu_5$ and $\mu_6$, which differ from the the KST expressions (due to
a typographical error in KST) by the substitutions
$\cbar\leftrightarrow\dbar$.

Finally we wish to give an explicit expression for the
transformation that relates a fundamental representation of the
symmetrizer $S ^0_{\alpha\beta}$ with the general representation $S
_{\alpha\beta}$.  We have seen in Sec.~\ref{s:symmetric} that a
symmetrizer $S _{\alpha\beta}$ is determined by a set of constants
$B_A$ and $C_A$, and similarly the fundamental representation $S
^0_{\alpha\beta}$ is determined by constants $B^0_A$ and $C^0_A$.  The
general symmetrizer is related to the fundamental by
Eq.~\eqref{e:stransform}, $S _{\alpha\beta}= T ^{-1\mu}{}_\alpha
S ^0_{\mu\nu}T ^{-1\nu}{}_\beta$.  This transformation can be
represented more concretely as relationships between the constants $B_A$
and $C_A$ that define $S _{\alpha\beta}$ and the constants $B^0_A$ and
$C^0_A$ that define $S^0_{\alpha\beta}$.  These relations are:
\begin{eqnarray}
B_1&=&(1+3\zbar)^2B_1^0,\label{e:b1}\\
B_2&=&B_2^0,\\
C_1&=&(\kbar+\ebar)^2C_1^0,\label{e:c1}\\
C_2&=&(\kbar-\half\ebar)^2 C_2^0,\\
C_3&=&{\cal A}^2 C_3^0 + {\cal B}^2 C_4^0 + 2 {\cal AB}C_5^0,\\
C_4&=&{\cal D}^2 C_3^0 + {\cal E}^2 C_4^0 + 2 {\cal DE}C_5^0,\\
C_5&=&{\cal AD} C_3^0 + {\cal BE} C_4^0 + ({\cal AE+BD})C_5^0,
\label{e:c5}
\end{eqnarray}
where
\begin{eqnarray}
{\cal A}&=&\kbar+3\abar+\cbar,\\
{\cal B}&=&\half\ebar+\abar+2\cbar,\\
{\cal D}&=&\ebar+3\bbar+\dbar,\\
{\cal E}&=&\kbar+\half\ebar+\bbar+2\dbar.
\end{eqnarray}
It is straightforward to verify that these $B_A$ and $C_A$ satisfy the
positivity conditions needed to guarantee that $S _{\alpha\beta}$ is
positive definite, so long as the $B^0_A$ and $C^0_A$ also satisfy those
conditions.


\end{document}